\documentclass{acmart}
\usepackage{hyperref}
\hypersetup{hidelinks=true}
\usepackage{amsmath,amsfonts}
\PassOptionsToPackage{hyphens}{url} 
\usepackage{graphicx,color} 
\usepackage{textcomp}
\usepackage{soul} 
\usepackage{booktabs, nicematrix} 
\usepackage{array}
\usepackage{multirow}
\usepackage{subfig}
\usepackage{paralist} 
\usepackage{threeparttable} 
\usepackage{tabularray}
\usepackage{tablefootnote}
\usepackage[inline]{enumitem} 
\usepackage{sistyle} 
\SIthousandsep{,}
\usepackage{url}
\usepackage{tcolorbox}
\usepackage{pdflscape}
\usepackage{adjustbox}
\usepackage{booktabs}
\usepackage{xcolor}
\usepackage{color, colortbl}
\usepackage{makecell}
\usepackage[font=small,skip=0pt]{caption}

\usepackage{algorithmic}
\usepackage{algorithm,algorithmic}
\def\BibTeX{{\rm B\kern-.05em{\sc i\kern-.025em b}\kern-.08em
    T\kern-.1667em\lower.7ex\hbox{E}\kern-.125emX}}
\AtBeginDocument{\definecolor{tmlcncolor}{cmyk}{0.93,0.59,0.15,0.02}\definecolor{NavyBlue}{RGB}{0,86,125}}

\newcommand{\quotes}[1]{``#1''} 

\newcommand{\autorefappendix}[1]{\hyperref[#1]{Appendix~\ref*{#1}}}  

\newcommand*\gray{\cellcolor{gray!25}}
\begin{document}

\title{Detection and Impact of Debit/Credit Card Fraud: Victims' Experiences}

\author{Eman Alashwali}
\affiliation{%
  \institution{King Abdulaziz University (KAU) and King Abdullah University of Science and Technology (KAUST)}
  \country{Saudi Arabia}
  \authornote{Eman Alashwali was a Collaborating Visitor at CMU while working on this paper. Both Eman and Ragashree contributed equally.}
  }
\email{ealashwali@kau.edu.sa}

\author{Ragashree Mysuru Chandrashekar}
\affiliation{%
  \institution{Carnegie Mellon University (CMU)}
  \country{United States}
  }
\email{ragashreeshekar@gmail.com}

\author{Mandy Lanyon}
\affiliation{%
  \institution{Carnegie Mellon University (CMU)}
  \country{United States}
  }
\email{mandy@cmu.edu}

\author{Lorrie Faith Cranor}
\affiliation{%
  \institution{Carnegie Mellon University (CMU)}
  \country{United States}
  }
\email{lorrie@cmu.edu}
\begin{tcolorbox}
This document is the author's manuscript for a paper to appear in Proceedings of the European Symposium on Usable Security (EuroUSEC), 2024.
\end{tcolorbox}

\begin{abstract}
It might be intuitive to expect that small or reimbursed financial loss resulting from credit or debit card fraud would have low or no financial impact on victims. However, little is known about the extent to which financial fraud impacts victims psychologically, how victims detect the fraud, which detection methods are most efficient, and how the fraud detection and reporting processes can be improved. To answer these questions, we conducted a 150-participant survey of debit/credit card fraud victims in the US. Our results show that significantly more participants reported that they were impacted psychologically than financially. However, we found no relationship between the amount of direct financial loss and psychological impact, suggesting that people are at risk of being psychologically impacted regardless of the amount lost to fraud. Despite the fact that bank or card issuer notifications were related to faster detection of fraud, more participants reported detecting the fraud after reviewing their card or account statements rather than from notifications. This suggests that notifications may be underutilized. Finally, we provide a set of recommendations distilled from victims' experiences to improve the debit/credit card fraud detection and reporting processes.
\end{abstract}
\maketitle
\section{Introduction}
Financial fraud represents a serious challenge globally to both individuals and society. Notably, identity theft fraud, a type of fraud that includes credit and debit card fraud according to the United States Federal Trade Commission (FTC) definitions, is the top-ranked reported type of fraud in 2021 and 2022 in the US, representing 23.43\% and 20.47\% of the reported frauds respectively~\cite{ftc23}. 
In 2022, the top three payment methods used in the reported frauds were credit cards, debit cards, and payment apps or services, with \$219.9 million, \$195.2 million, and \$163.5 million losses respectively~\cite{ftc23_2}. In 2021, according to a report on victims of identity theft by the Bureau of Justice Statistics (BJS), which also considers credit card and bank frauds under identity theft, 9\% of the US population aged 16 or older had experienced identity theft in the past 12 months, with nearly 4\% experiencing credit card fraud, and 3\% experiencing bank account fraud. The identity theft experiences did not only affect victims financially, but also resulted in 10\% of victims feeling extremely distressed, and 56\% spending a mean time of four hours to resolve the financial or credit issues~\cite{harrell23}.

In the US, debit/credit card fraud victims are legally protected by laws and regulations, such as the Electronic Fund Transfer Act (Regulation E)~\cite{ftc_2_24}. This limits victims' liability for unauthorized transfers if they give timely notice to the financial institution. For example, consumer liability is limited to \$50 if a notice is given to the financial institution within two business days~\cite{cfpb24}. Intuitively, this should limit the financial impact on victims as long as they are reimbursed or compensated for financial loss in a timely manner. While most fraud reports focus on financial loss, little attention is paid to understanding victims' experiences, such as how they detected the fraud and the psychological impact resulting from the fraud.   

This paper aims to bridge the gap in understanding victims' experiences with fraud. To this end, we surveyed 150 participants in the US who experienced a debit/credit card fraud incident within the last three years and asked them to report on their most recent fraud experience. 

This paper explores how victims detect fraud and to what extent fraud impacts victims psychologically. We also explore the relationship between psychological impact and the amount of financial loss. Finally, we use our data to provide recommendations for improving the fraud detection and reporting processes.

The key findings of this study are as follows: 
\begin{itemize}[noitemsep,nolistsep] %
    \item While notifications were related to faster fraud detection, fewer participants reported that they detected the fraud through notifications than through reviewing card or account statements, suggesting that notifications may be underutilized.
    \item Significantly more participants reported that they were psychologically impacted than financially impacted by fraud.
    \item Among our participants, the psychological impact was not related to the amount of financial loss, suggesting that people are at risk of being psychologically impacted regardless of the amount lost to fraud. 
\end{itemize}

Based on our findings, we suggest that banks and card issuers: utilize basic notifications by default; provide victims with explanations about how the fraud happened (if applicable) with actionable advice on how to protect themselves from future fraud; and examine solutions to remediate psychological impacts of fraud.

\section{Background and Related Work}
In this section, we provide a summary of related work on debit/credit card fraud detection techniques~(\autoref{sec:detect}). We then summarize some challenges faced by victims in reporting financial fraud~(\autoref{sec:report}). Finally, we summarize related work on the impact of financial fraud on victims~(\autoref{sec:impact}).

\subsection{Fraud Detection} \label{sec:detect}
As fraudulent transactions often share common features, automated systems can be trained to detect them~\cite{rymantubb_18}. Generally speaking, before approving each transaction, financial institutions use fraud detection systems that employ a classifier to determine whether the requested transaction is suspicious or not~\cite{rymantubb_18}. Since fraud techniques evolve over time, fraud detection systems also need to learn continuously~\cite{rymantubb_18}. 

Researchers have proposed a plethora of automated financial fraud detection methods that use Artificial Intelligence (AI) and Machine Learning (ML)~\cite{srivastava08,delamaire09,rymantubb_18,khatri20,seera21,ting24}. However, Ryman-Tubb et al. evaluated 51 AI and ML published methods for payment card fraud detection and found that only 8 have performance applicable for deployment in the industry~\cite{rymantubb_18}. The technical details of financial fraud detection methods are beyond the scope of this paper.

Despite numerous published research papers on automated financial fraud detection techniques, we have observed that many victims of financial fraud, including some of this paper's authors, detected fraudulent transactions through self-realization when they found transactions they did not make appeared in their bank statements. For example, Jansen and Luekfeldt analyzed 600 cases of phishing and malware banking fraud in the Netherlands~\cite{jansen_15}. They found that participants played a major role in discovering the fraud, after which they contacted their bank~\cite{jansen_15}. This included cases where victims had uncomfortable or suspicious feelings regarding an interaction, consulted family or friends, noticed unfamiliar payments or changes in their balance, or could not log in to their online bank account~\cite{jansen_15}. 

Very little research has investigated how victims of financial fraud discovered that they have been defrauded. In our study, we investigate detection methods from victims' experiences (\autoref{sec:results_detection}).

\subsection{Fraud Reporting and Resolution} \label{sec:report}
Prior studies reported some obstacles victims face in reporting fraud incidents, which in some cases affected victims' attitudes towards reporting and led them to not report the incident at all. A common reporting issue victims may face is the lack of information regarding the responsible authority~\cite{razaq_21} and a confusing list of reporting channels~\cite{cross_16}. Victims have multiple reporting channels to choose from: the store or company's customer service, the bank, credit card company or other payment provider, law enforcement, and consumer protection organizations~\cite{morgan_17,deliema_17}. Razaq et al.'s study on mobile-based financial fraud in Pakistan found that 49\% of the victims did not report the incident at all~\cite{razaq_21}. Some of the documented issues that led to not reporting the fraud are: difficulties in reporting, unsatisfactory outcomes, lack of evidence or feeling responsible about the fraud, and embarrassment~\cite{razaq_21,deliema_17,button_14,cross_16}.
Prior work identified victims' needs to improve the reporting and resolution processes, such as: having clear reporting channels, feeling like they are being listened to, receiving an acknowledgment that they have been victims of a crime, and receiving access to professional support services to treat the psychological consequences~\cite{cross_16}. Button et al.'s survey on victims from the UK found that most victims need a single point of contact to report the fraud, sympathetic response, and be listened to~\cite{button_09}.

In our study, we identify recommendations to improve the reporting process (\autoref{sec:results_reporting}).

\subsection{Fraud Impact}  \label{sec:impact}
Prior research showed that fraud can severely impact victims and their families. Ganzini et al. compared data for victims of violent (e.g. robbery) and non-violent (financial fraud) crimes~\cite{ganzini90}. They found that anxiety and depression disorders were the most common psychological implications for both types of crimes~\cite{ganzini90}. Cross et. al. interviewed 80 participants from Australia who experienced online fraud with losses of \$10,000 or more in the past four years~\cite{cross_16}. They found that financial impact was affected by the amount of money lost and other life circumstances of the victim~\cite{cross_16}. However, most participants experienced significant emotional and psychological impacts, and some experienced impacts on their health and family relationships~\cite{cross_16}. For many victims, the impact was long-lasting and led to behavior change~\cite{cross_16}. DeLiema et al. examined the financial and psychological impact of identity theft victims from older adults (aged 65 or older) by analyzing data from 2000 self-reported victims~\cite{deliema_21}. Their results suggested that the resulting harm from identity theft goes beyond financial loss. While only 7\% suffered from \quotes{out-of-pocket costs} (costs that were not reimbursed), 34\% found the experience distressing~\cite{deliema_21}. They also found that the length of time the information was stolen before detecting the fraud and the time spent to resolve the issue were positively related to emotional distress~\cite{deliema_21}. Riek and B{\"o}hme surveyed 1242 victims of cybercrime from 6 European countries about 7 types of consumer-facing cybercrime~\cite{riek18}. They found that scams are the severest type of cybercrime~\cite{riek18}. Moreover, they found that non-financial losses such as time, including protection time, exceed out-of-pocket (non-reimbursed) losses in most countries in the study~\cite{riek18}.

Unlike Cross et al.'s study~\cite{cross_16}, which included victims with high-value financial loss, and DeLiema et al.'s study~\cite{deliema_21}, which included only older adult victims, our study included adult victims with any amount of loss from any age. Moreover, unlike prior work, our study provides both qualitative and quantitative insights and studies the relationship between financial and psychological impact in further depth (\autoref{sec:results_impact}).
\section{Methods}
In this section, we describe our methods. We first present the study's scope and definitions~(\autoref{sec:def}). We then describe the recruitment processes~(\autoref{sec:recruitment}),  survey design~(\autoref{sec:design}), analysis~(\autoref{sec:analysis}), and ethical considerations~(\autoref{sec:ethics}). Finally, we list some limitations~(\autoref{sec:limitations}).

\subsection{Scope and Definitions} \label{sec:def}
Our study's scope is victims' experiences in financial fraud pertaining to debit/credit cards. For the purpose of our survey, we created a general definition of debit/credit card fraud and presented it to participants. Our definition includes fraud incidents that occurred with or without the victim's participation in the fraudulent transaction. It also does not restrict the fraud technique apart from a debit/credit card involved (associated with either a personal or business account). Thus, it includes incidents resulting from individual financial fraud~\cite{beals_15}, such as those related to products and services (e.g. overcharging), as well as those including identity theft (e.g. stolen card details that resulted in fraudulent purchases). We defined \textbf{credit/debit card fraud} to participants as follows: an \quotes{incident where any amount of money was charged to your debit/credit card without your knowledge or consent.} This definition was embedded in the screening survey (Q.2 in~\autorefappendix{app:screening_survey}) that asked participants whether they had experienced a debit/credit card fraud incident.

Throughout the paper, we use the term \textbf{victim} to refer to the cardholder who experienced the fraud incident, and the term \textbf{fraudster} to refer to the adversaries behind the fraud execution. We use the term \textbf{fraud} to refer to debit/credit card fraud.

\subsection{Recruitment} \label{sec:recruitment}
We recruited 150 participants for our online survey. The sample size was motivated by balancing cost-effectiveness and acceptable statistical power for scientific research~\cite{serdar21}. A power analysis using the G*Power statistical tool~\cite{gpower24} for the Chi-square ($\chi^2$) test family with effect size w = 0.3 (medium), $\alpha$ err prob = 0.05, Power (1 - $\beta$ err prb) = 0.90, Df = 2 suggests that 150 participants is sufficient.

We recruited survey participants using Prolific~\cite{prolific23}, a research-oriented online participant recruitment platform. We used a screening survey to identify participants who had experienced a financial loss in a debit/credit card fraud incident within the last three years. In addition, we screened for participants with at least a 90\% approval rate on Prolific who were at least 18 years old, residing in the United States, and able to read and write in English. Those who met the screening criteria and provided at least a 20-character\footnote{As part of the screening, participants were not told the minimum number of characters required. We chose 20 characters according to our rough estimate of a minimal valid response length.} description of a fraud incident and how they discovered it were automatically approved and immediately invited to take part in the main survey. After participants completed the main survey, we manually reviewed open-ended responses to both the screening survey and the main survey and excluded participants who gave invalid responses on either survey. We chose to approve the screening survey automatically and do manual validation later so that participants could do both surveys in one session.

Of the 268 participants who completed the screening survey, 170 were eligible for the main survey, and 154 completed the main survey. We excluded four invalid survey responses: two invalid responses reported \$0 direct loss, conflicting with our requirement that participants have experienced financial loss \quotes{even if it was later compensated or reimbursed;} one participant provided a fraud description not related to debit/credit cards; and one participant's open-ended responses were of low quality, indicating inattention. We ended up with 150 completed responses included in our analysis.

We paid participants \$0.50 for completing the screening survey and \$3.75 for completing the main survey. We began recruiting survey participants on November 13, 2023 and ended recruitment the same day after we reached 154 responses. 

\subsection{Survey Design} \label{sec:design}
We conducted the screening survey and main survey using Qualtrics~\cite{qualtrics23}, an online survey platform. 
The screening survey questions and main survey questions are detailed in~\autorefappendix{app:survey}.

Our main survey had six sections. In the first section, we asked participants multiple-choice questions about the incident context, such as what type of card was involved in the fraud incident, the type of transactions they had been using the card for (online, in-person, or both), and details such as transaction frequency and how they typically performed transactions (e.g., using an app or website, digital wallet, ATM, point of sale machine, etc.). Next, we asked questions related to the detection of the fraud incident, including questions about what was the first thing that triggered their attention to discover the debit/credit card fraud. If detection was through an alert or notification, we asked them about the notification type, medium, and timing, as well as how helpful they found the notification, and asked them to explain in an open-ended response. Next, we asked a set of questions about whether and how they reported the fraud and sought or received compensation. We asked those who reported the incident multiple-choice questions about the reporting and compensation processes and an open-ended question about how their bank or card issuer could improve the support offered in the reporting and compensation processes. We asked those who did not report the incident an open-ended question about why they did not report it. In the fourth section, we asked participants to what extent the experience impacted them financially, psychologically, and on their level of trust in conducting financial transactions, both at the time of the incident and today (i.e. the time of filling the survey). In the fifth section, we asked questions about measures participants have taken to prevent future fraud. Finally, participants were asked a few demographic questions. 

Several items in our survey benefited from an IRB-approved unpublished 5-participant pilot interview study conducted by some of this paper's authors and others in 2023. Some of the indirect financial impact items in Q.27 were inspired by the FINRA survey~\cite{arc15} and a report by the BJS~\cite{bjs13}. The list of the psychological impacts in Q.28 was obtained from Agrafiotis et al's taxonomy of cyber harms (12 items from the psychological harm category)~\cite{agrafiotis_18}, and an additional 4 items from the FINRA survey~\cite{arc15}. 

Before launching our survey, we conducted two pilot surveys with 21 participants to help us further refine our questions. Pilot data is not included in our results.

\subsection{Analysis} \label{sec:analysis}
We used quantitative and qualitative methods to analyze our data. We analyzed quantitative results using descriptive statistics. To test significance, if the samples in the test were independent, we used the Chi-square test (denoted as $\chi^2$)~\cite{mchugh13}, given that no more than 20\% of the expected frequencies are less than 5 and none of the cells is less than 1 (otherwise Fisher's exact test should be used, which we did not need to)~\cite{kim17}. If the samples in the significance test were related (paired) in repeated measures, such as when we asked all our participants whether or not they were impacted from three different aspects: financially, psychologically, and trust, in this case, the Chi-square test cannot be used as it requires independent samples~\cite{mchugh13}. Thus, we used the Cochran Q test to identify whether there are significant differences between the three different scenarios~\cite{mangiafico16}. If the Cochran test p-value was significant, we performed pairwise tests using McNemar's test to identify which pairs have significant differences~\cite{mangiafico16_2}. For reliable McNemar tests, the sum of discordant cells (b + c; i.e. the shaded cells shown in~\autoref{tab:nemar_attime_fin_vs_psych} --~\autoref{tab:nemar_today_psych_vs_trust} in~\autorefappendix{app:more_results}) in the contingency tables should be at least 5, 10, or 25, a condition that was met in our data where no discordant cells sum to less than 25~\cite{mangiafico16_2}. We used an alpha level of .05 for all statistical tests. If the p-value is less than .001, we report it as \quotes{p-value < .001.} We performed corrections for multiple tests using the Bonferroni method. All tests were computed using the R statistical tool~\cite{rproject24}. We computed the Chi-square effect size using Cramer's V (denoted as $\alpha_c$), Cochran effect size using Eta Square (denoted as $\eta_2$), and the MecNerma's effect size using Cohen's G (denoted as $|g|$). 

We grouped some answer choice categories together to facilitate analysis. Namely, for Q.16 on fraud detection speed, we offered participants 10 answer choices including \quotes{immediately,} \quotes{within a few minutes,} \quotes{within an hour,} up to \quotes{more than a month.} We grouped the first three categories as \quotes{fast} and the remainder as \quotes{not-fast.} We also grouped Q.26 categories which offered participants 12 direct financial loss choices starting from \quotes{\$0.01 to \$10} to \quotes{more than \$10,000} into two categories \quotes{small,} and \quotes{not-small.} Finally, we grouped the fraud impact categories for Q.29 and Q.30, combining \quotes{strongly impacted me negatively} and \quotes{somewhat impacted me negatively} into one \quotes{impacted} category to contrast with \quotes{did not impacted me negatively at all.} The latter is the extreme opposite category to having any level of impact. We grouped these categories for analysis for reasons deemed appropriate to collapse categories~\cite{vaus14,dusen20}: to make them more relevant to our research problem, and to highlight patterns, especially as we combined relevant categories with low frequencies. The \quotes{strongly} impacted category has low frequencies in multiple measures (some measures have only one, six, and nine participants out of 150 participants). Also, our exploratory study's scope does not aim to investigate the impact severity levels, thus, treating the two impact categories (i.e. \quotes{strongly} and \quotes{somewhat}) separately will not add new information to our findings. It is worth noting that the \quotes{strongly} impacted data has a similar pattern to that of \quotes{somewhat} impacted (see the red and yellow color bars in~\autoref{fig:impact}). That is, more participants reported they were psychologically and trust impacted than financially impacted. Thus, combining the \quotes{strongly} and \quotes{somewhat} impacted categories allowed us to highlight those patterns without introducing new relationships~\cite{vaus14}.

To qualitatively analyze the open-ended responses, two researchers used Template Analysis, a style of thematic analysis that combines both inductive and deductive coding, with emphasis on hierarchical coding without specific prescription regarding the number of levels required and what levels represent~\cite{king23, brooks14}. After the surveys were completed, both researchers familiarized themselves with the data by reading through the responses. The first researcher created an initial codebook (the template) and coded all the data. Then, the second researcher reviewed the codes applied by the first researcher. Both researchers met over multiple sessions to discuss any disagreements in the coding and adjusted the codebook (added, updated, and deleted codes) until they reached a satisfactory version of the codebook agreeable to both researchers. The codebook is provided in~\autorefappendix{app:codebook}.

\subsection{Ethical considerations} \label{sec:ethics}
We obtained approval from the CMU Institutional Review Board (IRB). All participants were presented with an online consent form at the beginning of the screening survey. We informed participants about the risk that some questions may bring unpleasant memories to some individuals. We also informed them that they have the right to opt out at any point during the survey.

\subsection{Limitations} \label{sec:limitations}
First, we used Prolific to recruit participants and Prolific workers are known to be more technically skilled than the general population. However, they are reasonably diverse, and Prolific has been shown to provide more generalizable results than other popular recruitment platforms~\cite{tang22}. Second, as we asked participants to self-report the most recent fraud incident that occurred within the last three years, responses might be prone to recall bias. However, this seems to have limited impact on our study as over than half of our participants (54\%) reported that their fraud experience occurred \quotes{within the last year} or \quotes{within the last few months.}\footnote{The full breakdown of the data is as follows: within the last: few months (17.33\%); year (36.67\%); 2 years (25.33\%); 3 years (20.67\%).} Third, the incident descriptions are limited to participants' understanding of what happened. We used self-reported data as we otherwise had no access to incident reports and our main interest was in participants' experiences and reflections on the incidents. Fourth, we acknowledge potential misinterpretation of one of the survey questions (Q.16) regarding the approximate length of time it took the participant to detect the fraud. The question asked about detection from the time the fraudulent transaction occurred but we suspect that some participants who detected the fraud when reviewing the account or card statement might have answered with respect to the time they reviewed the statement as we see more than expected short time frame for discovering fraud from reviewing statement. However, the implications of this potential misinterpretation are limited and would not change the conclusion (explained in further details in the results~\autoref{sec:method_and_fast}). Fifth, our study is focused on the US context, includes only US-based participants, and is framed around US financial regulations and practices. Regulations as well as bank and card issuer practices differ between jurisdictions, which needs to be considered when interpreting our results outside the US context. Sixth, our sample is limited to 150 participants. While this sample size is sufficient for our exploratory study that aims to identify high-level patterns and trends, a larger sample with sufficient data in scarce categories (such as large direct financial loss) might reveal relationships that our study could not reveal. Finally, our study is exploratory and does not measure nuanced levels of severity of impact. It also does not measure relationships between impact and income, or unreimbursed (a.k.a out-of-pocket) loss. Moreover, it does not capture pre-incident security behavior.

\section{Results} \label{sec:results}
In what follows we report our results. Where applicable, we denote participants as P, followed by the participant ID (e.g. P\_001). We provide frequencies of each qualitative theme to provide a sense of our data and highlight the most/least common themes, but these numbers should not be interpreted as generalizable results. Finally, references to survey question numbers are for the main survey in~\autorefappendix{app:main_survey}, unless stated otherwise.

\subsection{Participants}
Out of the 150 participants, 78 (52.0\%)  identified themselves as female, 71 (47.3\%) as male, and one as non-binary. Their ages ranged from 18 to \quotes{75 or older,} with a majority falling between 25–34 (42.7\%) and 35-44 (22.7\%). A total of 111 (74.0\%) were employed, 9 (6.0\%) were students, and the remaining were distributed among several other statuses. When asked about the name of the bank or card issuer, participants reported a wide variety of banks and card issuers, with 65 (43.3\%) reporting one of the following major US banks as their card issuer: Bank of America, Capital One, Chase, Citi, and Wells Fargo. See~\autoref{tab:demographics} in~\autorefappendix{app:more_results} for further demographic data.

Almost all participants 149 (99.3\%) reported that the fraud occurred on a personal account, while only one reported a business account. 89 (59.3\%) reported that fraud occurred on a debit card, 58 (38.7\%) on a credit card, and only 3 (2.0\%) on a combination debit/credit card.~\autoref{tab:financialbehaviour} in~\autorefappendix{app:more_results} provides further details on participants' use of the card involved in the fraud incident they described. 
 
\subsection{Fraud Incidents}
We asked participants to describe their most recent debit/credit card fraud experience and how they discovered the fraud in an open-ended question (Q.6 \& Q.7 in the screening survey in~\autorefappendix{app:screening_survey}). Our qualitative analysis shows that only 6 (4.0\%) participants mentioned incidents that involved the possession of the original physical card by the fraudster, such as lost or stolen cards, and one participant mentioned a skimmed card. This indicates that most of the reported incidents 
were card-not-present fraud, which includes online, phone, or mail transactions~\cite{egan23}. These results are in line with recent reports regarding the rise of card-not-present fraud in the US~\cite{lebow23,egan23}. 

We analyzed participants' descriptions of the incident to determine the fraudulent transaction mode. 53 (35.3\%) of participants' answers indicated online transactions, such as \quotes{Someone used my credit card for an Amazon order} (P\_001). Only 19 (12.7\%) participants indicated offline transaction mode, which we identified from mentions of specific locations, such as \quotes{Unexpected charges showed up for pizza in an area where I do not reside} (P\_021), or other cues such as fraud that involved cash withdrawals. The remaining participants did not specify the fraudulent transaction mode. 

We also analyzed their descriptions to determine the type of fraudulent transaction. 87 (58.0\%) participants indicated unauthorized purchases. Participants also mentioned other types of unauthorized transactions, such as overcharging (4), bank identity theft (3), unauthorized withdrawal (2), never-received items (1), and unauthorized check issuance (1). The remaining participants (54) did not specify the fraudulent transaction type, mentioning only that an unauthorized transaction was made: \quotes{There were online charges on my credit card that I didn't make} (P\_013). Participants' descriptions of the fraud included mentions of shopping or stores in general (22), food and beverages (19), electronics (7), gas (7), subscription or membership (6), clothes (4), entertainment (3), transportation (3), money-laundering-like website (2), and porn (1).

Location was another significant detail mentioned by participants, with a total of 34 (22.7\%) participants reporting the fraudulent activity occurred outside their geographic location including a different state (23), continent (4), area (3), city (2), or country (2). 

Fourteen individuals indicated a compromise of a non-bank online account that contained the debit/credit card details, such as \quotes{Someone hacked my walmart account and used my card that was on there} (P\_016). In one case, a participant mentioned, \quotes{My family took money out of my account} (P\_050). 

\subsection{Fraud Detection} \label{sec:results_detection}
In this section, we summarize the reported method of detection that led participants to discover the fraud~(\autoref{sec:detection_method}). Then, we summarize the type and medium of notifications the participants received~(\autoref{sec:detection_type_medium}). Finally, we analyze the impact of the detection method (checking card or account statement vs. receiving a notification) on the detection time~(\autoref{sec:method_and_fast}).

\subsubsection{Method of Detection} \label{sec:detection_method}
We asked participants to identify the first thing that triggered their attention to discover their debit/credit card fraud (Q.9 in~\autorefappendix{app:main_survey}). As shown in~\autoref{tab:detection_method}, nearly 89\% of participants detected the fraud through one of three mechanisms: checking their card or account statement and finding transactions they did not make (50.0\%), receiving a form of alert or notification from their bank or card issuer (34.7\%), or receiving a form of alert or notification from a third-party (4.0\%).

\begin{table}[tbp!]
\renewcommand{\arraystretch}{1.25}
\caption{The methods through which participants first discovered that they had been defrauded, ordered from the most to least reported method.}
\label{tab:detection_method}
\centering
\footnotesize
\begin{tabular}{p{7cm}|ll}
    \toprule
    \multicolumn{3}{c}{\textit{N = 150}} \\
    \midrule
    Method of detection & \# & \% \\
    \midrule
        I checked my card or account statement and found a transaction(s) I did not make & 75 & (50.0\%) \\
    
        I received a form of alert or notification from my bank or card issuer (e.g. SMS, phone call, letter, email, etc. for withdrawal or fraud notifications) & 52 & (34.7\%) \\

        Other        & 10 & (6.7\%) \\

        I received a form of alert or notification from a third-party (e.g. SMS, phone call, letter, email, etc. for withdrawal or fraud notifications) (*please specify what type of third-party): & 6 & (4.0\%) \\
        
        I realized I might have fallen for a scam or something did not seem right, which led me to do further checks and subsequently find the fraud & 5 & (3.3\%) \\

        I checked my card or account statement and found a transaction(s) that I made, but went to an unintended recipient(s) & 1 & (0.7\%) \\
        
        I learned about the fraud from family, friends, or other people, which led me to do further checks and subsequently find the fraud & 1 & (0.7\%) \\

        I learned about the fraud from public channels (e.g. social media), which led me to do further checks and subsequently find the fraud & 0 & (0.0\%) \\
      \bottomrule
\end{tabular}
\end{table}

\subsubsection{Type and Medium of Notification} \label{sec:detection_type_medium}
Out of the 58 participants who detected the fraud through a form of notification, a total of 45/58 (77.6\%) reported that the notification type was either \quotes{Suspicious transaction} or \quotes{Fraudulent transaction.} Only a few participants mentioned other specific types of notifications, including  \quotes{Withdrawal} (5), \quotes{Account balance} (1), and \quotes{International transaction} (1). See~\autoref{tab:notif_type} for a full list of the reported notification types.

\begin{table}[tbp!]
\footnotesize
\centering
\caption{List of reported notification types by the participants, ordered from the most to least reported type.}
\label{tab:notif_type}
    \begin{tabular}{l|ll}
    \toprule 
    \multicolumn{3}{c}{\textit{N = 58}}\\
    \midrule
    Notification Type             &  \# &  \% \\ 
    \midrule
    Suspicious transaction        & 36                & (62.1\%) \\ 
    Fraudulent transaction        & 9                 &  (15.5\%)\\ 
    Withdrawal                    & 5                 & (8.6\%)\\
    Other                         & 4                 & (6.9\%)\\
    I cannot remember             & 2                 & (3.4\%)\\ 
    Account balance               & 1                 & (1.7\%)\\
    International transaction     & 1                 & (1.7\%)\\
    Low balance                   & 0                 & (0.0\%)\\
    \bottomrule
    \end{tabular}
\end{table}

Short Messaging Service (SMS) was the most reported notification medium with 32/58 (55.2\%), followed by email at 21/58 (36.2\%), and phone call at 14/58 (24.1\%). 20 participants reported receiving the notification through multiple channels. See~\autoref{tab:notif_medium} for a full breakdown of the reported notification mediums.

\begin{table}[tbp!]
\footnotesize
\centering
\caption{List of reported notification mediums by the participants, ordered from the most to least reported medium.}
\label{tab:notif_medium}
    \begin{tabular}{p{6cm}|ll}
    \toprule
    \multicolumn{3}{c}{\textit{N = 58}}\\
    \midrule
    \makecell[l]{Notification Medium (multiple answers)}          &  \# & \%  \\ 
    \midrule 
    Short text message (SMS)  &   32            &  (55.2\%)\\ 
    Email                     &   21           &  (36.2\%)\\ 
    Phone call                &   14            &  (24.1\%)\\
    Push notification through the bank or card issuer's app or Internet banking website 
                              &   8            & (13.8\%)\\ 
    
    Message through the bank or card issuer's app or Internet banking website
                                                &    6            &  (10.3\%)\\ 
   Automated voice message   &   4             &  (6.9\%)\\ 
     
    Other                     &   1            &  (1.7\%)\\ 
    Post letter               &   0            &  (0.0\%)\\ 
    \bottomrule
    \end{tabular}
\end{table}
Almost all the participants who received a form of notification from their bank or card issuer across all types of notifications (57/58) reported they found the notification helpful in detecting the fraudulent activity (Q.14 in~\autorefappendix{app:main_survey}). 51 (87.9\%) said that the notifications were \quotes{Extremely helpful,} 5 (8.6\%) said \quotes{Somewhat helpful,} one \quotes{Slightly helpful,} and only one said, \quotes{Not at all helpful.}

\subsubsection{Impact of Detection Method on Detection Time} \label{sec:method_and_fast}
We now look closer at the top two most reported detection methods, reviewing card or account statement and receiving notifications, to see which method is related to faster fraud detection. For each detection method, we analyzed participants' responses to our question about how long it took them to realize that they had been defrauded from the time the fraudulent transaction occurred (Q.16 in~\autorefappendix{app:main_survey}). If the participant answered \quotes{Immediately,} \quotes{Within a few minutes,} or \quotes{Within an hour,} we classified this as \quotes{fast} detection, otherwise, we classified it as \quotes{Not fast.} Participants who answered \quotes{I cannot remember} (1 participant from the bank statement detection method and 5 from the notifications detection method) were excluded from this analysis. Our data originally had a total of 76 participants who reported fraud detection by reviewing their bank or card statement and a total of 58 through receiving a form of alert or notification (see~\autoref{tab:detection_method}). This exclusion left us with a total of 75 participants who reported statement, and 53 participants who reported notifications as detection methods, respectively. 

Our results show that notifications are more related to faster detection of the fraud, where 37/53 (69.8\%) of the reported fraud incidents from the notifications group were detected within an hour or less, compared to 29/75 (38.7\%) of incidents detected from the statement group. The difference is statistically significant ($\chi^2$ = 12.061, p-value < .001, $\alpha_c$ = 0.291). However, despite the fact that notifications led to faster detection, our results show that more participants detected the fraud through statements (53 via notifications compared to 75 via statement). It should be noted that the number of participants who reported fast fraud detection through their card or account statement may be inflated due to potential misinterpretation of Q.16. We suspect that some participants provided the time from the time they checked their statement instead of what we asked in the question \quotes{From the time your debit/credit card fraudulent transaction occurred} as it seems unlikely that a lot of people would happen to check their account statement within an hour of a fraudulent charge being made. We observed such confusion in our pilot studies, and therefore reworded this question placing the key part first, and using bold and italic font to emphasize the key parts of this question. Despite those changes to the question, more than expected number of participants reported fast detection through statements. We suspect some confusion about this question remains and the number of participants who reported fast detection through statements is inflated. Having said that, this potential confusion only suggests that the difference between notifications and statements in terms of enabling fast fraud detection might be larger than what our results indicate. The difference is already statistically significant and having it even larger difference would not change the conclusion. That is, in either case, notifications are related to faster detection of the fraud than reviewing the bank or card statement, and the difference is statistically significant.

\subsubsection{How Alerting and Detection Can Be Improved} 
We asked participants to provide suggestions for improving the alerting and detection capabilities (Q.18 in~\autorefappendix{app:main_survey}). Some of the suggestions were related to the alert type participants wanted to receive. For example, 11 participants suggested preventive notifications 
such as alerts for suspicious or fraudulent transactions, 11 suggested alerts for unusual behavior, and 12 participants suggested alerts to approve or decline transactions. As P\_109 explained: \quotes{maybe like a quick text or call from the bank to make sure it's really me making a purchase.} 16 participants suggested alerts for transactions originating from different geographical locations (2 of them specifically mentioned international transactions), alerts for any purchase transactions (8), large transactions (6), transactions over a certain limit (1), and alerts for repeated purchases within a short duration of time (1). Five participants would like to receive alerts for any transaction, one participant suggested alerts related to security activities including change of PIN, password, etc., and 2 participants recommended enabling alerts by default. With respect to alert medium, SMS was the most suggested form of delivery (15), followed by email (7), phone call (5), and through the app (2). Additionally, 6 participants preferred receiving alerts through multiple channels. Many participants mentioned speed, with 11 participants stating they wanted alerts to be sent immediately. Two participants emphasized the importance of the legitimacy of these texts, without which they may be misinterpreted as scams. 

Many of these requested alert capabilities are already offered by some US banks, although they may not be offered by default (they need to be configured by the cardholder).


\subsection{Fraud Reporting and Compensation} \label{sec:results_reporting}
In this section, we summarize participants' responses about their experiences reporting the fraud~(\autoref{sec:reporting}), and what they believe can improve the reporting process~(\autoref{sec:explanation}).

\subsubsection{Reporting and Not Reporting} \label{sec:reporting}
While under-reporting of financial fraud was found to be a problem in some countries such as Pakistan as reported in Razaq et al's study~\cite{razaq_21}, which is mostly related to socio-cultural and regulation issues, it does not seem to be an issue with our US participants. The majority of our participants 129 (86.0\%) said they reported the fraud to their bank or card issuer. Among those who reported the incident, the majority 102/129 (79.1\%) sought compensation, and 109 (84.5\%) of those who reported the incident said they were fully compensated. We asked those who did not report the fraud about the reasons for not doing so (Q.20 in~\autorefappendix{app:main_survey}). Almost all of them explained that the bank or card issuer detected the fraud for them as P\_40 explained: \quotes{It was the bank that notified me. So that was unnecessary,} which eliminated the need for reporting. Two participants said that the loss was not important enough as P\_074 explained: \quotes{I only lost a few dollars because the other transactions were declined,} and in the case of P\_050 where she described that the fraudster was a family member said: \quotes{Did not want to get my family in trouble.}

We then asked participants for suggestions to improve the reporting and compensation processes (Q.25 in~\autorefappendix{app:main_survey}). We found that speeding up the reporting and/or compensation processes was one of the key points mentioned by 22 participants. On the other hand, those who did not have any suggestions and expressed their satisfaction with their experience, appreciated the fast process overall (11).

\subsubsection{Explaining What Happened} \label{sec:explanation}
We asked participants who said they reported the fraud, whether the bank provided them with an explanation of the fraud incident (Q.23 in~\autorefappendix{app:main_survey}). Only 22/129 (17.1\%) said they received a full explanation, while the remaining majority either received no explanation at all 63/129 (48.8\%), or a partial explanation 41/129 (31.8\%). 

Providing explanations and additional information about the fraud incident was a key point mentioned by 26 participants when asked for suggestions to improve the reporting and compensation processes (Q.25 in~\autorefappendix{app:main_survey}). Participants wanted more information about how, what, where, and by whom the fraud occurred. In addition, some participants suggested banks or card issuers provide them with preventive tips \quotes{to prevent [this] from happening again} (P\_069). Lack of explanation can negatively impact some people as P\_045 explained: \quotes{The[re] was definatley some details left out that have me confused.}

\subsection{Fraud Impact} \label{sec:results_impact}
In this section, we summarize participants' responses about the fraud's direct, indirect, and psychological impact~(\autoref{sec:impact_type}). Then, we analyze the reported financial vs. psychological vs. trust impact to find which type of impact has impacted more participants~(\autoref{sec:fin_vs_psych_impact}). Finally, we analyze the relationship between the amount of direct financial loss and the different types of impacts~(\autoref{sec:fin_vs_psych_relation}). 

\subsubsection{Type of Impact} \label{sec:impact_type}
\paragraph{Direct Financial Impact.} \label{sec:direct_impact}
We asked participants to report the direct financial loss resulting from the fraud incident, which we described to them as \quotes{the amount the fraudster charged} to their card \quotes{even if it was later compensated or reimbursed} (Q.26 in~\autorefappendix{app:main_survey}). 83.3\% of the reported direct financial loss did not exceed \$500, suggesting that current detection techniques are reasonably helpful in limiting the amount of direct financial loss. See~\autoref{tab:direct_loss} for a full breakdown of the direct loss results.

\begin{table}[t!]
  \footnotesize
  \centering
  \caption{Direct financial losses reported by participants who experienced the fraud, ordered from small to large loss.}
  \label{tab:direct_loss}
  \begin{tabular}{l|ll}
      \toprule
      \multicolumn{3}{c}{\textit{N = 150}} \\
      \midrule
      Direct Financial Loss & \textbf{\#} & \textbf{\%} \\
      \toprule
        \$0 (no direct financial loss)  & 0 & (0.0\%)\\
        From \$0.01 to \$10 & 11 & (7.3\%)\\
        From \$11 to \$50 & 29 & (19.3\%)\\
        From \$51 to \$100 & 34 & (22.7\%)\\
        From \$101 to \$500 & 51 & (34.0\%)\\
        From \$501 to \$1,000 & 10 & (6.7\%)\\
        From \$1,001 to \$2,000 & 6 & (4.0\%)\\
        From \$2,001 to \$4,000 & 4 & (2.7\%)\\
        From \$4,001 to \$6,000 & 0 & (0.0\%)\\
        From \$6,001 to \$8,000 & 1 & (0.7\%)\\
        From \$8,001 to \$10,000 & 0 & (0.0\%)\\
        More than \$10,000 & 0 & (0.0\%)\\
        I cannot remember & 1 & (0.7\%)\\
        Other             & 3 & (2.0\%)\\
      \bottomrule
\end{tabular}
\end{table}

\paragraph{Indirect Financial Impact.} \label{sec:indirect_impact}
We then asked participants to report the indirect financial losses they had experienced. Participants selected one or more of eleven common indirect financial impacts or \quotes{Other,} or the exclusive \quotes{None of these.} A considerable number of participants 67/150 (44.7\%) reported they did not experience any indirect financial loss. However, \quotes{Emotional distress} was selected by 69 (46.0\%), followed by experiencing \quotes{Loss of trust from others} by 15 (10.0\%), suggesting that emotional and psychological impacts are the top-ranked indirect losses, which we examine in further detail next. Issues such as \quotes{Difficulty obtaining loans or credit} were reported only by a few participants, while more serious indirect financial losses such as \quotes{Loss of job due to the fraud incident} were reported by none. The full list of reported indirect financial impacts is provided in~\autoref{tab:indirect_loss}. 

\begin{table}[t!]
  \footnotesize
  \centering
  \caption{Indirect losses reported by participants who experienced the fraud, ordered from the most to least reported impact.}
  \label{tab:indirect_loss}
  \begin{tabular}{p{6cm}|ll}
      \toprule
      \multicolumn{3}{c}{\textit{N = 150}} \\
      \midrule
      Indirect Loss (multiple answers) & \textbf{\#} & \textbf{\%} \\
      \midrule
        Emotional distress & 69 & (46.0\%) \\
        None of these & 67 & (44.7\%) \\
        Loss of trust from others & 15 & (10.0\%) \\
        Costs of credit monitoring or identity theft protection services & 11 & (7.3\%) \\
        Loss of income due to the fraud incident & 8 & (5.3\%) \\
        Other         & 6 & (4.0\%) \\
        Damaged personal reputation & 3 & (2.0\%) \\
        Negative impact on credit score or higher interest rates for borrowing & 3 & (2.0\%)\\
        Increased insurance premiums or difficulty obtaining insurance coverage & 1 & (0.7\%) \\
        Difficulty obtaining loans or credit & 1 & (0.7\%) \\
        Costs of therapy or counseling & 0 & (0.0\%) \\
        Loss of job due to the fraud incident & 0 & (0.0\%) \\
        Legal and attorney fees for pursuing legal action or defending against accusations & 0 & (0.0\%) \\
      \bottomrule
  \end{tabular}
  \end{table}

\paragraph{Psychological Impact.} \label{sec:psych_impact}
To gain more insights, we asked participants to report all the psychological negative impacts they experienced from a list of 16 psychological impacts or provide their answer in the \quotes{Other} choice. 136/150 (90.7\%) participants reported one or more negative psychological impacts, with an average of 3.8 impacts (excluding those who reported \quotes{None of these}) per participant. The most reported impacts were: \quotes{Stress} 103 (68.67\%), \quotes{Worry or anxiety} 79 (52.7\%), \quotes{Frustration} 78 (52.0\%), and \quotes{Feeling upset} 73 (48.67\%). Sometimes people were worried about others besides themselves, as P\_060 added: \quotes{I was worried, not only for my account, but several others who are family and elderly. I had to spend time notifying people that their accounts might be compromised as well.} The full list of reported impacts is provided in~\autoref{tab:psych_impact}.
 \begin{table}[t!]
  \footnotesize
  \centering
  \caption{List of psychological negative impacts reported by participants who experienced the fraud, ordered from the most to least reported impact.}
  \label{tab:psych_impact}
  \begin{tabular}{p{5cm}|ll}
    \toprule
    \multicolumn{3}{c}{\textit{N = 150}} \\
    \midrule
    Psychological impact (multiple answers) & \textbf{\#} & \textbf{\%} \\
    \midrule
        Stress & 103 & (68.7\%) \\
        Worry or anxiety & 79 & (52.7\%) \\
        Frustration & 78 & (52.0\%) \\
        Feeling upset & 73 & (48.7\%) \\
        Confusion & 36 & (24.0\%) \\
        Discomfort & 31 & (20.7\%) \\
        Feeling unsafe & 28 & (18.7\%) \\
        Difficulty in trusting others & 25 & (16.7\%) \\
        Guilt & 15 & (10.0\%) \\
        None of these & 14 & (9.3\%) \\
        Difficulty in sleeping & 12 & (8.0\%) \\
        Low satisfaction & 10 & (6.7\%) \\
        Negative changes in perception & 10 & (6.7\%) \\
        Shame & 8 & (5.3\%) \\
        Loss of self-confidence & 5 & (3.3\%) \\
        Other         & 3 & (2.0\%) \\
        Depression & 0 & (0.0\%) \\
        Embarrassment & 0 & (0.0\%) \\
    \bottomrule
  \end{tabular}
  \end{table}

\subsubsection{Financial vs. Psychological vs. Trust Impact} \label{sec:fin_vs_psych_impact}
\paragraph{Size of Financial, Psychological, and Trust Impacts.}
To measure the financial, psychological, and trust impacts on victims, we asked participants to rank how the fraud incident impacted them financially, psychologically, and on their level of trust in performing financial transactions. We asked them to rank those 3 impacts twice: with respect to the time of the incident and today (the time of filling out the survey). Participants were presented with 3 choices (Q.29 and Q.30 in~\autorefappendix{app:main_survey}) as follows: \quotes{Strongly impacted me negatively,} \quotes{Somewhat impacted me negatively,} and \quotes{Did not impacted me negatively at all.} 

\begin{figure*}[t!]
\centering
\resizebox{\textwidth}{!}{%
\begin{minipage}[b]{\textwidth}
    \centering
    \begin{minipage}[b]{0.7\textwidth}
        \subfloat[Impact at the time of the incident.\label{fig:time_impact}]{%
        \includegraphics[width=\textwidth]{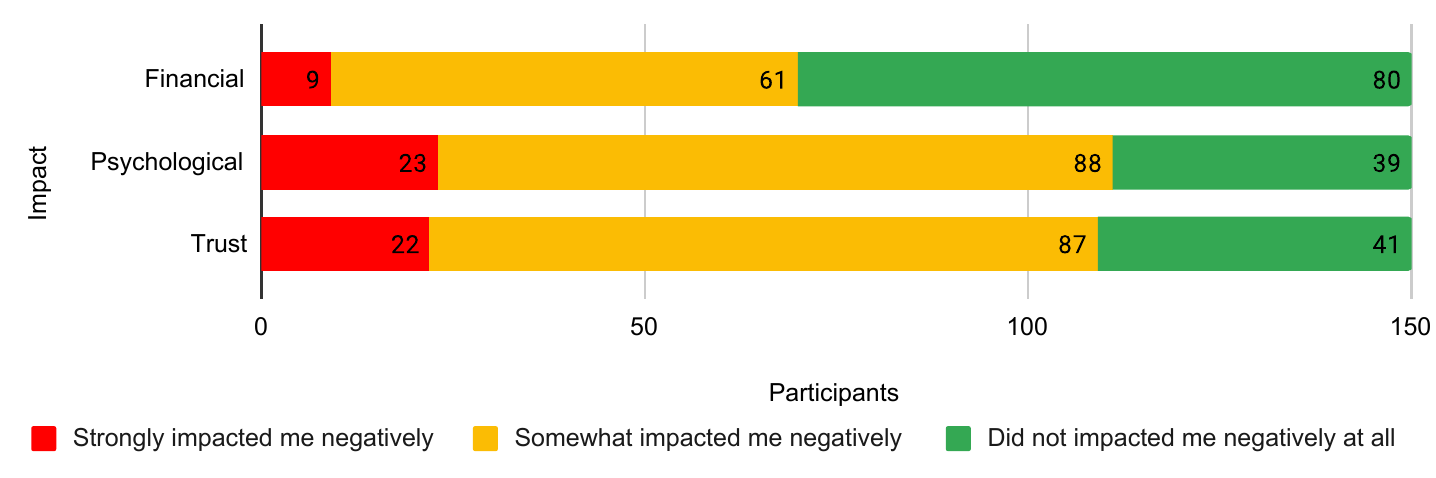}
        }
    \end{minipage}
    \vfill
    \begin{minipage}[b]{0.7\textwidth}
        \subfloat[Impact today.\label{fig:now_impact}]{%
        \includegraphics[width=\textwidth]{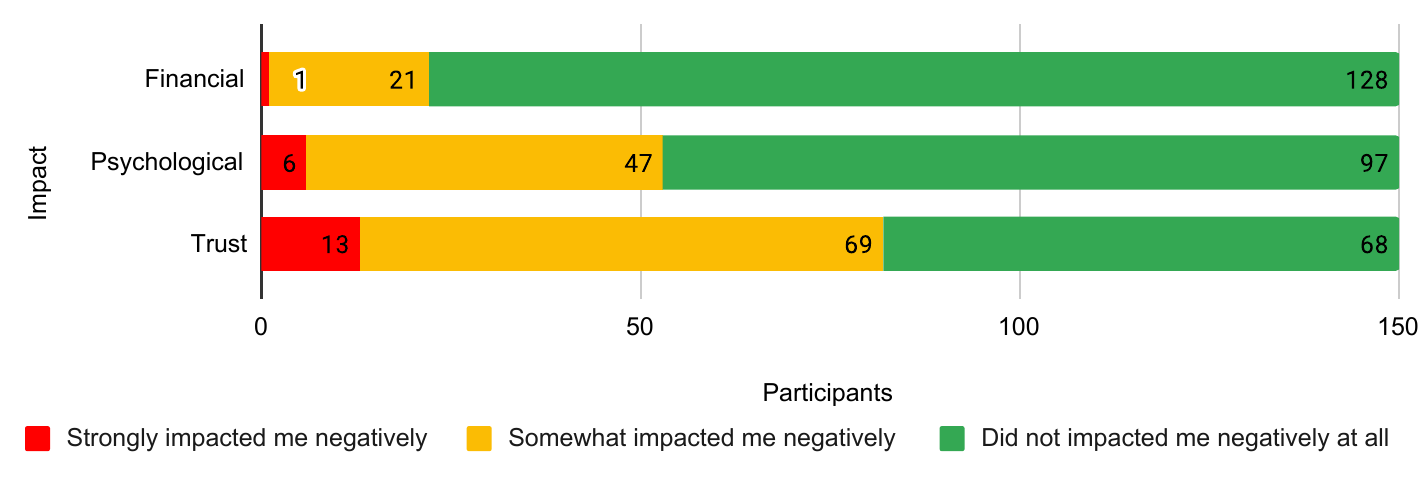}
        }
    \end{minipage}
\end{minipage}
}%
\caption{Financial vs. psychological vs. trust impact on victims with respect to the time of the incident and today.} 
\label{fig:impact}
\end{figure*}

As~\autoref{tab:fin_vs_psyc} and~\autoref{fig:impact} show, more participants reported they were psychologically and trust impacted than financially impacted. This is for both questions: with respect to the time of the incident and today. Moreover, while the financial and psychological impacts of the fraud reduced over time, the loss of trust in performing financial transactions did not reduce as much.
\begin{table}[t!]
\footnotesize
\centering
\caption{Number of participants who reported financial, psychological, and trust impacts at the time of the incident and today.}
\label{tab:fin_vs_psyc}
\begin{tabular}{l|ll|ll|ll}
\toprule
    \multicolumn{7}{c}{\textit{N = 150}} \\
    \midrule 
      & \makecell{\# Fin.\\impacted} & \% & \makecell{\# Psych.\\ impacted} & \% &  \makecell{\# Trust\\impacted} & \% \\ 
    \hline
     At time   & 70/150   & (46.7\%) &   111/150  & (74.0\%) & 109/150 & (72.7\%) \\  
     \hline
     Today      & 22/150 & (14.7\%)   &    53/150 & (35.3\%)  &  82/150 & (54.7\%) \\ 
    \bottomrule
\end{tabular}
\end{table}
\par To test whether there are significant differences between the number of participants who were impacted financially, psychologically, and on their trust in performing financial transactions, we coded participants who reported \quotes{strongly} or \quotes{somewhat} impacted as \quotes{impacted,} and  those who reported they were not impacted at all as \quotes{not impacted.} We then performed the Cochran tests. We found significant differences between the three types of impacts at both points in time: at the time of the incident ($Q$ = 38.166, p-value < .001, $\eta_2$ = 0.127) and today ($Q$ = 74, p-value < .001 $\eta_2$ = 0.246). 

Then, to identify which pair of impacts significantly differ, we conducted pairwise McNemar tests for each of the possible pair of impacts as follows: (financial vs. psychological), (financial vs. trust), and (psychological vs. trust).
\begin{table}[t!]
    \footnotesize
    \centering
    \caption{Pairwise McNemar's tests for the financial vs. psychological vs. trust impacts. The p-values were adjusted using the Bonferroni method. The effect size for McNemar tests was computed using Cohen's G ($|g|$).}
    \label{tab:fin_vs_psych_vs_trust}
    \begin{tabular}{l|l|l|l}
    \toprule
    \multicolumn{4}{c}{\textit{N = 150}} \\
    \midrule
    Impacted (at time)  &  \makecell[l]{McNemar's\\$\chi^2$}    &  \makecell{Adjusted\\p-value}     &   $|g|$   \\
    \hline 
    Fin. vs. Psych.     & 25.09         & \cellcolor{gray!30}< .001   & 0.306 \\
    Fin. vs. Trust      & 23.4         & \cellcolor{gray!30}< .001  & 0.3  \\
    Psych. vs. Trust    & 0.111           & 1.000                       & 0.028 \\                    
    \midrule 
    Impacted (today)    &  $\chi^2$    &  \makecell{Adjusted\\p-value}    &   $|g|$    \\
    \hline 
    Fin. vs. Psych.     & 24.641      & \cellcolor{gray!30}< .001  & 0.397 \\
    Fin. vs. Trust      & 56.25       & \cellcolor{gray!30}< .001  & 0.469  \\
    Psych. vs. Trust    & 19.558      & \cellcolor{gray!30}< .001  & 0.337  \\  
    \bottomrule
    \end{tabular}
\end{table}
As detailed in~\autoref{tab:fin_vs_psyc} and~\autoref{tab:fin_vs_psych_vs_trust}, our results suggest that, at both points in time (at the time of the incident and today), more participants reported they were psychologically and trust impacted than financially impacted, and these differences are significant. 

\paragraph{Relationship Between the Amount of Financial Loss and Financial, Psychological, and Trust Impacts.} \label{sec:fin_vs_psych_relation}
We now look at whether there is a relationship between the amount of direct financial loss (Q.26 in~\autorefappendix{app:main_survey}) and the financial, psychological, and trust impacts. To this end, we classified participants into two categories based on the reported amount of direct financial loss: small loss of \$100 or less coded as \quotes{small,} and larger losses (more than \$100) coded as \quotes{not small.} We excluded participants who selected \quotes{I cannot remember} (1) or \quotes{Other} (3) for the direct financial loss question (Q.26). This left us with 146 participants included in this section's analysis. Similar to previous sections, we coded participants who reported being \quotes{strongly} or \quotes{somewhat} impacted as \quotes{impacted,} and those who reported they were not impacted at all as \quotes{not impacted.} We ended up with~\autoref{tab:fin_loss_vs_psych} listing the direct financial loss categories, the number of participants whose reported direct financial loss fit in that category, and the number of participants who reported they were financially, psychologically, and trust impacted in each direct financial loss category. Note that in this test we only consider impact with respect to the time of the incident (Q.29 in~\autorefappendix{app:main_survey}). 

\begin{table*}[t!]
\footnotesize
\centering
\caption{Columns 2--4 respectively list the total number of participants who reported \quotes{small} vs. \quotes{not small} direct financial loss; the number of participants in the corresponding direct loss group who reported financial; psychological; and trust impacts (at the time of the incident).}
\label{tab:fin_loss_vs_psych}
\begin{tabular}{l|c|ll|ll|ll}
\toprule
    \multicolumn{8}{c}{\textit{N = 146}} \\
    \midrule 
    Direct loss in \$ & \makecell[l]{\# experienced\\loss in \$} & \makecell[l]{\# Fin.\\impacted} & \%  & \makecell[l]{\#Psych.\\impacted} & \% & \makecell[l]{\#Trust\\impacted} & \%  \\
    \hline
    \makecell[l]{Small\\(\$0.01 to \$100)} & 74 & 33/74 & (44.6\%)  & 52/74 & (70.3\%)  & 52/74 & (70.3\%) \\
    \hline
    \makecell[l]{Not small\\(more than \$100)} & 72 & 36/72 & (50\%)  & 56/72 & (77.8\%)  & 53/72 & (73.6\%)   \\ 
\bottomrule
\end{tabular}
\end{table*}

We then used the Chi-square test for the following pairs of variables: (direct financial loss vs. financial impact), (direct financial loss vs. psychological impact), and (direct financial loss vs. trust impact).

Our Chi-square tests suggest that there is no relationship between the amount of direct financial loss and: the financial, psychological, and trust impact. That is, financial loss vs. financial impact ($\chi^2$ = 0.427, p-value = 0.513, $\alpha_c$ = 0.040), financial loss vs. psychological impact ($\chi^2$ = 1.068, p-value = 0.301, $\alpha_c$ = 0.069), and financial loss vs. trust impact ($\chi^2$ = 0.201, p-value = 0.653, $\alpha_c$ = 0.021).

\paragraph{Impact of Fraud on Behavior.}
As a result of the fraud experience, 144 (96.0\%) participants reported they had taken one or more security measures to prevent future fraud on their card (Q.31). As~\autoref{tab:measures} details, the majority of participants primarily resorted to manual measures in their control to safeguard themselves, such as regularly checking their account or card statement, reported by 104 (69.3\%) participants, and exercising caution and vigilance when conducting financial transactions, reported by 93 (62.0\%) participants. However, more automated (smarter) techniques such as notifications and security configurations were mentioned at a lower scale. For example, 57 (38.0\%) participants reported they set up alerting mechanisms for their account or card, 55 (36.7\%) reported implementing security measures from the bank or card issuer, 26 (17.3\%) reported implementing security measures from third parties, while 10 (6.7\%) said they changed their bank or card issuer.

\begin{table*}[t!]
    \renewcommand{\arraystretch}{1.25}
    \centering
    \footnotesize
    \caption{List of measures participants reported had taken to prevent future fraud on their debit/credit card, ordered from the most to least reported measure.}
    \label{tab:measures}
    \begin{tabular}{l|ll}
        \toprule
         \multicolumn{3}{c}{\textit{N = 150}} \\
         \midrule 
         Measure taken (multiple answers) & \# & \% \\
         \hline 
         I now regularly check my account or card statement                                                      & 104  & (69.3\%) \\
         I am now more cautious and vigilant when conducting financial transactions	                            & 93   & (62.0\%) \\
         I have set up alerting mechanisms for my account or card	                                            & 57   & (38.0\%) \\
         I have implemented security measures from the bank or card issuer to protect my financial transactions	& 55   & (36.7\%) \\
         I have implemented security measures from third-parties to protect my financial transactions	        & 26   & (17.3\%) \\
         I switched my bank or card issuer	                                                                    & 10   & (6.7\%) \\
         I did not undertake any measures	                                                                    & 6	   & (4.0\%) \\
         Other (please specify)	                                                                                & 6	   & (4.0\%) \\
         \bottomrule
    \end{tabular}
\end{table*}

\section{Discussion}
Our discussion is limited to our study's scope on victims' experiences, which is an important source to understand the credit/debit card fraud landscape, what consumers are experiencing, and their needs. We do not analyze what financial institutions are implementing due to a lack of such data, which is a common limitation in fraud research.

\paragraph{Notifications: Efficient, Yet Underutilized.}
Our results show that notifications are related to faster detection of fraud than reviewing the card or account statement. This might be intuitive, however, our results show that more participants reported that they detected the fraud through checking their card or account statement than through notifications, suggesting that notifications are underutilized. It is our experience that not all US banks enable notifications (e.g. for any transaction) by default. However, when we asked participants about suggestions for banks to improve their alerting and detection capabilities many participants suggested a wide variety of notifications with the SMS as the most mentioned medium. 

We recommend that well-tailored basic notifications, such as any transaction notifications or approve or decline card-not-present transactions, are best suited to be enabled by default, with the option to opt out. We acknowledge the trade-off in enabling basic notifications by default: they may incur additional costs to banks, may annoy some customers, or customers may become habituated to them such that they do not notice them. The literature has investigated the habituation to warnings and notifications~\cite{anderson14,anderson16,vance19}. However, Reeder et. al.'s study on browser warnings suggests that contextual factors affect users' decisions toward adherence or bypassing warnings, and that habituation has a smaller effect on users' decision-making process than previously thought~\cite{reeder18}. Other studies also suggested that perceived disruption of mobile notifications is influenced by contextual factors, such as the sender-receiver relationship, the importance of the information the notification contains, and personality traits~\cite{mehrotra16,sahami14,fischer10}. Mehrotra et al. found that users tend to click highly disruptive notifications if the content is important information~\cite{mehrotra16}. Future work might explore notifications in the banking context including experimental studies on user perceptions and behaviors towards notifications in the banking context, which notifications should be enabled by default, and the most usable and secure delivery medium. 

\paragraph{Explanation as a Preventive Mechanism.}
While most banks stress the importance of customer awareness to combat fraud, nearly half of our participants did not receive any explanation at all from their banks or card issuers regarding the fraud. When we asked participants for suggestions to improve the support provided to them in the reporting and compensation processes, our qualitative data shows that explanation was the most mentioned theme. Several participants particularly pointed out the need for an explanation to \quotes{avoid this happening again in the future} (P\_148). 

We recommend that banks or card issuers provide an explanation of the fraud incident to victims when applicable. Although card fraud is sometimes completely beyond the victim's control, some incidents could be prevented if victims were more aware of how to protect themselves from phishing, card skimmers, and other types of attacks. In addition, knowing that incidents can happen beyond their control may help motivate people to turn on notifications or be more vigilant about checking their statements for fraudulent transactions. We acknowledge that banks and card issuers may not always be able to provide an explanation of the fraud incident for various reasons, such as lack of information, cost of investigation, or security. However, card issuers may still be able to provide actionable and more tailored advice to help victims avoid future fraud. 

\paragraph{The Forgotten Psychological Impact.}
Our results suggested that the psychological impact affects more participants than the financial impact, regardless of the amount of financial loss. Most of our participants were reimbursed for the direct financial loss, yet reported that they remain psychologically impacted months or years later. This suggests that reimbursement alone is not sufficient to remediate the psychological impact. It is unclear what banks and consumer protection organizations are doing to address the psychological impact on financial fraud victims. 

We suggest that banks and consumer protection organizations examine and further research solutions to remediate the psychological impact. Offering professional psychological support for the psychologically impacted victims is a solution worth examining to remediate the issue. In addition, explanations may provide some assurance to victims, helping them to feel better prepared to protect themselves in the future. Moreover, future work could study the impact, harms, and potential benefits of psychological support in greater detail.

\paragraph{Behavioral Analytics to Combat Credit/Debit Card Fraud.}
Behavioral analytics, which can predict user behavior based on analyzing data from their past behavior, has long been associated with targeted advertisements in web and mobile applications. In the context of advertisements, behavioral analytics tend to be viewed as privacy-invasive techniques~\cite{ur12,turow09}, and users face difficulties in opting out of online advertisements~\cite{leon12}. However, in the banking context, ideally, banks and their customers have a form of trust relationship, and banks already audit customers' transactions. Our qualitative data shows that when customers see a value (security) in return for behavioral analytics of their financial behavior, behavioral analytics are viewed as a welcomed feature. Our qualitative results show that many participants would like their banks or card issuers to notify them if they identified out-of-ordinary behavior. While most banks may already use forms of behavioral analytics, many participants' answers indicate that they were not aware of them.

\section{Conclusion}
In this study, we surveyed 150 victims of credit/debit card fraud in the US to report on their most recent experiences. Our results show that the psychological impact affected more participants than the financial impact, regardless of the amount of financial loss. Among our participants, the psychological impact was not related to the amount of financial loss, suggesting that people are at risk of being psychologically impacted regardless of the amount lost to fraud. While notifications resulted in faster detection of fraud than checking the bank statement, they were less utilized in detecting the fraud.
\section*{Acknowledgements}
Eman Alashwali acknowledges the financial support of the Ibn Rushd Program at King Abdullah University of Science and Technology (KAUST). This work was funded in part by the Innovators Network Foundation. We thank Prof. Marc Dacier from KAUST for feedback on earlier versions of this paper; Jenny Tang, Elijah Bouma-Sims, and Eric Zeng for helpful discussions on statistical analysis; Talal Ali, David Mberingabo, Yiming Zhong, and Pauline Mevs for their contributions to the initial pilot interview study during the Usable Privacy and Security course at CMU.


\bibliographystyle{plain}
\bibliography{ref}
\appendix
\clearpage
\twocolumn
\section{Results Tables} \label{app:more_results}
\begin{table}[h!]
\caption{Participants' general demographics.}
\label{tab:demographics}
\centering
\footnotesize
\begin{tabular}{p{4cm}|ll}
    \toprule
    \multicolumn{3}{c}{\textit{N = 150}} \\
    \midrule
    Gender & \# & \% \\
    \hline
        \quad Male & 71 & (47.3\%)\\
        \quad Female & 78 & (52.0\%)\\
        \quad Non-binary & 1 & (0.7\%)\\
        \quad Prefer to self describe & 0 & (0.0\%)\\
        \quad Prefer not to answer & 0 & (0.0\%)\\
    \hline
    Age & \# & \% \\
    \hline
        \quad 18 to 24  & 19 & (12.7\%)\\
        \quad 25 to 34 & 64 & (42.7\%)\\
        \quad 35 to 44  & 34 & (22.7\%)\\
        \quad 45 to 54  & 17 & (11.3\%)\\
        \quad 55 to 64 & 10 & (6.7\%)\\
        \quad 65 to 74 & 4 & (2.7\%)\\
        \quad 75 or older & 2 & (1.3\%)\\
        \quad Prefer not to answer & 0 & (0.0\%)\\
    \hline
    Education & \# & \% \\
    \hline
        \quad Doctorate & 2 & (1.3\%)\\
        \quad Master degree & 19 & (12.7\%)\\
        \quad Bachelor degree & 60 & (40.0\%)\\
        \quad Associate degree & 23 & (15.3\%)\\
        \quad High school diploma or GED & 43 & (28.7\%)\\
        \quad Less than High school degree & 1 & (0.7\%)\\
        \quad Other                        & 2 & (1.3\%)\\
    \hline
    Current Occupation & \# & \% \\
    \hline
        \quad Student & 9 & (6.0\%)\\
        \quad Full-time employee & 81 & (54.0\%)\\
        \quad Part-time employee & 21 & (14.0\%)\\
        \quad Self-employed or business owner & 9 & (6.0\%)\\
        \quad Full-time homemaker & 4 & (2.7\%)\\
        \quad Unemployed, and looking for a job & 12 & (8.0\%)\\
        \quad Unemployed, and not looking for a job & 2 & (1.3\%)\\
        \quad Unable to work & 4 & (2.7\%)\\
        \quad Retired & 4 & (2.7\%)\\
        \quad Other   & 4 & (2.7\%)\\
    \hline
    Income & \# & \% \\
    \hline
        \quad Less than \$20,000 & 15 & (10.0\%)\\
        \quad \$20,000 to \$39,999 & 19 & (12.7\%)\\
        \quad \$40,000 to \$59,999 & 32 & (21.3\%)\\
        \quad \$60,000 to \$79,999 & 22 & (14.7\%)\\
        \quad \$80,000 to \$99,999 & 27 & (18.0\%)\\
        \quad \$100,000 to \$149,999 & 18 & (12.0\%)\\
        \quad \$150,000 or more & 13 & (8.7\%)\\
        \quad Prefer not to answer & 4 & (2.7\%)\\
    \hline
    CS, IS, IT or CE degree & \# & \% \\
    \hline
        \quad Yes & 18 & (12.0\%)\\
        \quad No & 132 & (88.0\%)\\
    \hline 
    Cybersecurity Degree & \# & \% \\
    \hline
        \quad Yes & 1 & (0.7\%)\\
        \quad No & 149 & (99.3\%)\\
    \bottomrule
\end{tabular}
\end{table}

\begin{table}[h!]
\caption{Details provided by participants about their use of the debit- or credit-card involved in the fraud incident they describe in the survey.}
\label{tab:financialbehaviour}
\centering
\footnotesize
\begin{tabular}{p{5cm}|ll}
    \toprule
    \multicolumn{3}{c}{\textit{N = 150}} \\
    \midrule
    Account Type & \# & \% \\
        \midrule
            \quad Personal Account  & 149 & (99.3\%)\\
            \quad Business Account  & 1 & (0.7\%)\\
            \quad I cannot remember & 0 & (0.0\%)\\
            \quad Other             & 0 & (0.0\%)\\
        \midrule
    Type of Card & \# & \% \\
    \midrule
            \quad Debit Card & 89 & (59.3\%)\\
            \quad Credit Card & 58 & (38.7\%)\\
            \quad Debit and credit card in one card  & 3 & (2.0\%)\\
            \quad I cannot remember & 0 & (0.0\%)\\
    \midrule
    Category & \# & \% \\
        \midrule
            \quad Bank Visa or Mastercard debit/credit card & 131 & (87.3\%)\\
            \quad Discover or American Express debit/credit card & 6 & (4.0\%)\\
            \quad Co-branded debit/credit card  & 8 & (5.3\%)\\
            \quad I cannot remember & 1 & (0.7\%)\\
            \quad I don’t know & 2 & (1.3\%)\\
            \quad Other        & 2 & (1.3\%)\\
        \midrule
    Modes of Transactions (non-exclusive) & \# & \% \\
        \midrule
            \quad Online & 121 & (80.7\%)\\
            \quad In person & 111 & (74.0\%)\\
            \quad I cannot remember & 6 & (4.0\%)\\
        \midrule
    Avenues of Online Transactions (non-exclusive) & \# & \% \\
        \midrule
            \quad Apps and websites & 113/121 & (93.4\%)\\
            \quad The bank or card issuer's app or website & 47/121 & (38.8\%)\\
            \quad Digital wallets & 47/121 & (38.8\%)\\
            \quad Third-party financial apps & 64/121 & (52.9\%)\\
            \quad I cannot remember & 3/121 & (2.5\%)\\
            \quad Other             & 2/121 & (1.7\%)\\
        \midrule
    Frequency of Online Transactions  & \# & \% \\
        \midrule
            \quad At least once a day & 11/121 & (9.1\%)\\
            \quad At least once every few days & 45/121 & (37.2\%)\\
            \quad At least once a week & 30/121 & (24.8\%)\\
            \quad At least once every few weeks & 21/121 & (17.4\%)\\
            \quad At least once a month & 9/121 & (7.4\%)\\
            \quad At least once every few months & 2/121 & (1.7\%)\\
            \quad At least once every six months & 0/121 & (0.0\%)\\
            \quad At least once a year & 1/121& (0.8\%)\\
            \quad Less than once a year & 2/121 & (1.7\%)\\
            \quad Never & 0/121 & (0.0\%)\\
            \quad I cannot remember & 0/121 & (0.0\%)\\
        \midrule
    Avenues of In-Person Transactions (non-exclusive) & \# & \% \\
        \midrule
            \quad Automated Teller Machines & 43/111 & (38.7\%)\\
            \quad Point of sale machine & 101/111 & (91.0\%)\\
            \quad I cannot remember & 1/111 & (0.9\%)\\
            \quad Other             & 2/111 & (1.8\%)\\
        \midrule
    Frequency of In-Person Transactions & \# & \% \\
        \midrule
            \quad At least once a day & 18/111 & (16.2\%)\\
            \quad At least once every few days & 42/111 & (37.8\%)\\
            \quad At least once a week & 24/111 & (21.6\%)\\
            \quad At least once every few weeks & 11/111 & (9.9\%)\\
            \quad At least once a month & 2/111 & (1.8\%)\\
            \quad At least once every few months & 8/111 & (7.2\%)\\
            \quad At least once every six months & 2/111 & (1.8\%)\\
            \quad At least once a year & 1/111 & (0.9\%)\\
            \quad Less than once a year & 2/111 & (1.8\%)\\
            \quad Never & 0/111 & (0.0\%)\\
            \quad I cannot remember & 1/111 & (0.9\%)\\
        \bottomrule  
\end{tabular}
\end{table}
 
\begin{table*}[h!]
\caption{Financial vs. psychological impact -- \textbf{at the time of the incident.} The sum of all cells equals \textit{N} = 150.}
\label{tab:nemar_attime_fin_vs_psych}
\centering
\footnotesize
\begin{tabular}{l|l|l}
    \toprule
      & Psych. impacted & Psych. not-impacted  \\
      \hline
       Fin. impacted & 57 & \gray 13  \\
       \hline 
       Fin. not-impacted & \gray 54 & 26 \\
    \bottomrule  
\end{tabular}
\end{table*}

\begin{table*}[h!]
\caption{Financial vs. trust impact -- \textbf{at the time of the incident.} The sum of all cells equals \textit{N} = 150.}
\label{tab:nemar_attime_fin_vs_trust}
\centering
\footnotesize
\begin{tabular}{l|l|l}
    \toprule
      & Trust impacted & Trust not-impacted  \\
      \hline
       Fin. impacted & 57 & \gray 13  \\
       \hline 
       Fin. not-impacted & \gray 52 & 28 \\
    \bottomrule  
\end{tabular}
\end{table*}

\begin{table*}[h!]
\caption{Psychological vs. trust impact -- \textbf{at the time of the incident.} The sum of all cells equals \textit{N} = 150.}
\label{tab:nemar_attime_psych_vs_trsut}
\centering
\footnotesize
\begin{tabular}{l|l|l}
    \toprule
      & Trust impacted & Trust not-impacted  \\
      \hline
       Psych. impacted & 92 & \gray 19 \\
       \hline 
       Psych. not-impacted & \gray 17 & 22 \\
    \bottomrule  
\end{tabular}
\end{table*}

\begin{table*}[h!]
\caption{Financial vs. psychological impact -- \textbf{today.} The sum of all cells equals \textit{N} = 150.}
\label{tab:nemar_today_fin_vs_psych}
\centering
\footnotesize
\begin{tabular}{l|l|l}
    \toprule
      & Psych. impacted & Psych. not-impacted  \\
      \hline
       Fin. impacted & 18 & \gray 4  \\
       \hline 
       Fin. not-impacted & \gray 35 & 93  \\
    \bottomrule  
\end{tabular}
\end{table*}

\begin{table*}[h!]
\caption{Financial vs. trust impact -- \textbf{today.} The sum of all cells equals \textit{N} = 150.}
\label{tab:nemar_today_fin_vs_trust}
\centering
\footnotesize
\begin{tabular}{l|l|l}
    \toprule
      & Trust impacted & Trust not-impacted  \\
      \hline
       Fin. impacted & 20 & \gray 2   \\
       \hline 
       Fin. not-impacted & \gray  62 & 66   \\
    \bottomrule  
\end{tabular}
\end{table*}

\begin{table*}[h!]
\caption{Psychological vs. trust impact -- \textbf{today.} The sum of all cells equals \textit{N} = 150.}
\label{tab:nemar_today_psych_vs_trust}
\centering
\footnotesize
\begin{tabular}{l|l|l}
    \toprule
      & Trust impacted & Trust not-impacted  \\
      \hline
       Psych. impacted & 46 & \gray 7  \\
       \hline 
       Psych. not-impacted & \gray 36  & 61  \\
    \bottomrule  
\end{tabular}
\end{table*}
\clearpage
\twocolumn
\section{Survey}\label{app:survey}
\small
\autoref{app:recruitment_survey} lists the recruitment materials.~\autoref{app:screening_survey} lists the screening survey.~\autoref{app:main_survey} lists the main study's survey. Section headings and text between [square brackets] were not shown to participants. We added them here for clarity.

\subsection{Recruitment Materials} \label{app:recruitment_survey}

\noindent \textbf{Title}: Your debit/credit card fraud experiences

\noindent \textbf{Reward}: \$.50 (screening) \${3.75} (main) (approximately \${15}/hr)

\noindent \textbf{Estimated completion time}: {15} mins  

\noindent \textbf{Description}: Our research team at Carnegie Mellon University is searching for people to participate in a survey about their debit/credit card fraud experiences. Participants should have experienced debit/credit fraud to participate.

If you are interested in participating in our survey, please complete this initial screening survey, which should just take 2 minutes. You will be paid \$0.50 for completing the screening survey through Prolific.

If you are selected for the study’s main survey, you will be asked if you agree to be taken to that survey and then be directed to the main survey that will last about 15 minutes. You will be paid \$3.75 for completing this main survey through Prolific.

All responses to the survey will be kept confidential.

\noindent \textbf{Devices you can use to take this study}: Mobile, Tablet, Desktop
   
\subsection{Screening Survey} \label{app:screening_survey}
\noindent \textbf{Q.0:} Please enter your unique Prolific ID. \\
\noindent Please note that this response should auto-fill with the correct ID.\\
\noindent [free response field]

\noindent \textbf{Q.1:} Do you have a debit or credit card(s)?
\begin{itemize}[noitemsep,topsep=0pt]
    \item Yes
    \item No
\end{itemize}

\noindent [If response to Q.1 is \quotes{yes}]\\
\noindent \textbf{Q.2:} Have you ever experienced a debit/credit card fraud incident, where any amount of money was charged to your debit/credit card without your knowledge or consent? 
\begin{itemize}[noitemsep,topsep=0pt]
    \item Yes
    \item No   
\end{itemize}

\noindent [If Q.1 and Q.2 answers are \quotes{yes}] \\ 
\noindent \textbf{Q.3:} Approximately, when was your \textbf{most recent} experience of debit/credit card fraud?
\begin{itemize}[noitemsep,topsep=0pt]
    \item Within the last few months
    \item Within the last year
    \item Within the last 2 years
    \item Within the last 3 years
    \item Within the last 4 years
    \item Within the last 5 years
    \item More than 5 years ago
    \item I cannot remember 
\end{itemize}

\noindent [If Q.3 answer is \quotes{within the last 3 years} or less] \\
\noindent \textbf{Q.4:} In the past 3 years, approximately, how many times have you experienced debit/credit card fraud incidents?
\begin{itemize}[noitemsep,topsep=0pt]
    \item Once
    \item Twice
    \item Three times
    \item More than three times
    \item Other (please specify): [free response field]
\end{itemize}

\noindent [If Q.3 answer is \quotes{within the last 4 years} or more] \\
\noindent \textbf{Q.5:} In the past 5 years, approximately, how many times have you experienced debit/credit card fraud incidents?
\begin{itemize}[noitemsep,topsep=0pt]
    \item Once
    \item Twice
    \item Three times
    \item Four time
    \item Five times
    \item More than five times
    \item Other (please specify): [free response field]
\end{itemize}

\noindent \textbf{Q.6:} In your own words, please describe what happened in your \textbf{most recent} debit/credit card fraud experience? \\
\noindent [Free response field.]

\noindent \textbf{Q.7:} How did you discover the fraud in your \textbf{most recent} debit/credit card fraud experience? \\
\noindent [Free response field]

\noindent \textbf{Q.8:} What type of account was associated with the described debit/credit card fraud incident? 
\begin{itemize}[noitemsep,topsep=0pt]
    \item Personal account
    \item Business account
    \item I cannot remember
    \item Other (please specify): [free response field]
\end{itemize}

\noindent [If Q.8 answer is \quotes{personal} or \quotes{business} account] \\
\noindent \textbf{Q.9:} What was the country of origin of the bank or card issuer associated with the described debit/credit card in which you experienced the fraud incident?
\begin{itemize}[noitemsep,topsep=0pt]
    \item United States
    \item Other (please specify): [free response field]
\end{itemize}

\noindent [If: Q.1 answer is \quotes{yes}; Q.2 answer is \quotes{yes}; Q.3 answer is \quotes{within the last 3 years} or less; Q.6 and Q.7 answers are at least 20 characters long; and Q8 answer is personal or business account]

\noindent \textbf{Q.10:}Are you willing to participate in an online survey about your experience of debit/credit card fraud? (it takes around \textbf{15 min}. and you will be paid \textbf{\$3.75} for participation through Prolific) \newline 
\textbf{If you selected \quotes{Yes} you will be redirected to the survey now.} 
\begin{itemize}[noitemsep,topsep=0pt]
    \item Yes
    \item No   
\end{itemize}

\noindent[If Q.10 is \quotes{yes}, the following text is shown]\\
\noindent Please think only about the time of \textbf{your most recent experience of debit/credit card fraud you described previously in the screening survey} and answer all the following questions in that context.
 
\textbf{As a reminder, here is your answer for describing the fraud experience: }\\
\noindent[Q.6 answer is shown here] 

\textbf{Here is your answer for how you discovered it: \\}
\noindent[Q.7 answer is shown here]

\noindent [If Q.10 is \quotes{Yes}, automatic direction to the main survey]

\subsection{Main Survey} \label{app:main_survey}
\noindent [Section headings were not visible to participants.]

\noindent SECTION \#1: THE INCIDENT CONTEXT

\noindent \textbf{Q.1:} What type of card did you experience the debit/credit card fraud incident on?
\begin{itemize}[noitemsep,topsep=0pt]
    \item Debit card
    \item Credit card
    \item Debit and Credit card in one card (combination card)
    \item I cannot remember
\end{itemize}

\noindent \textbf{Q.2:} What type of debit/credit card did you experience the fraud incident on?
\begin{itemize}[noitemsep,topsep=0pt]
    \item Bank Visa or Mastercard debit/credit card
    \item Discover or American Express debit/credit card
    \item Co-branded debit/credit card (when organizations or companies such as retail stores or airlines partner with card issuer)
    \item I cannot remember
    \item I don’t know
    \item Other (please specify): [free response field]
\end{itemize}
  
\noindent \textbf{Q.3:} What was the name of the bank or card issuer associated with the described debit/credit card that you experienced the fraud incident on?\\
\noindent [Free response field]

\noindent\textbf{Q.4:} At the time of the incident, what type of transactions had you been using this debit/credit card for?  (select all that apply)
\begin{itemize}[noitemsep,topsep=0pt]
    \item Online
    \item In-person
    \item I cannot remember [exclusive]
\end{itemize}

\noindent [If Q.4 answer is \quotes{online}] \\
\noindent \textbf{Q.5:} At the time of the incident, which of the following had you been using to perform \textbf{online} transactions using this card? (select all that apply)
\begin{itemize}[noitemsep,topsep=0pt]
    \item Apps and websites (e.g. shopping apps/websites)
    \item The bank or card issuer’s app or website (e.g. to transfer money)
    \item Digital wallets (e.g. Apple Pay and Google Pay)
    \item Third-party financial apps (Paypal, Venmo, etc.)
    \item I cannot remember [exclusive]
    \item Other (please specify): [free response field]
\end{itemize}

\noindent [If Q.4 answer is \quotes{online}] \\
\noindent \textbf{Q.6:} At the time of the incident, approximately how often had you been using this card for \textbf{online} transactions?
\begin{itemize}[noitemsep,topsep=0pt]
    \item At least once a day
    \item At least once every few days
    \item At least once a week
    \item At least once every few weeks
    \item At least once a month
    \item At least once every few months
    \item At least once every six months
    \item At least once a year
    \item Less than once a year 
    \item Never
    \item I cannot remember 
\end{itemize}

\noindent [If Q.4 answer is \quotes{in-person}] \\
\noindent \textbf{Q.7:} At the time of the incident, which of the following had you been using to perform \textbf{in-person} transactions using this card? (select all that apply)
\begin{itemize}[noitemsep,topsep=0pt]
    \item Automated Teller Machines (e.g. ATMs)
    \item Point of sale machine (e.g. those found in shops, restaurants, etc.)
    \item I cannot remember [exclusive]
    \item Other (please specify): [free response field]
\end{itemize}

\noindent [If Q.4 answer is \quotes{in-person}] \\
\noindent \textbf{Q.8:} At the time of the incident, approximately, how often had you been using this card for \textbf{in-person }transactions?
\begin{itemize}[noitemsep,topsep=0pt]
    \item At least once a day
    \item At least once every few days
    \item At least once a week
    \item At least once every few weeks
    \item At least once a month
    \item At least once every few months
    \item At least once every six months
    \item At least once a year
    \item Less than once a year 
    \item Never
    \item I cannot remember
\end{itemize}~\\

\noindent SECTION \#2: DETECTION (POST-FRAUD) \\
\noindent Please think only about the time of \textbf{your most recent experience of debit/credit card fraud that you described previously} in this survey, and answer all the following questions in that context.

\noindent \textbf{Q.9:} Which of the following was \ul{\textbf{the first thing}} that triggered your attention to discover your debit/credit card fraud? 
\begin{itemize}[noitemsep,topsep=0pt]
    \item I checked my card or account statement and found a transaction(s) I did not make
    \item I checked my card or account statement and found a transaction(s) that I made, but went to an unintended recipient(s)
    \item I realized I might have fallen for a scam or  something did not seem right, which led me to do further checks and subsequently find the fraud
    \item I received a form of alert or notification \textbf{from my bank or card issuer} (e.g. SMS, phone call, letter, email, etc. for withdrawal or fraud notifications)
    \item I received a form of alert or notification\textbf{ from a third-party} (e.g. SMS, phone call, letter, email, etc. for withdrawal or fraud notifications) (*please specify what type of third-party): [free response field] 
    \item I learned about the fraud from public channels (e.g. social media), which led me to do further checks and subsequently find the fraud
    \item I learned about the fraud from family, friends, or other people, which led me to do further checks and subsequently find the fraud
    \item Other (please specify): [free response field]
\end{itemize}

\noindent [If Q.9 answer is \quotes{I received a form of alert or notification from my bank or card issuer...} or \quotes{I received a form of alert or notification from a third-party...}] \\
\noindent \textbf{Q.10:} What was the type of the received alert or notification?
\begin{itemize}[noitemsep,topsep=0pt]
    \item Withdrawal
    \item Low balance
    \item Account balance
    \item Suspicious transaction
    \item Fraudulent transaction
    \item International transaction
    \item I cannot remember 
    \item Other (please specify): [free response field]
\end{itemize}

\noindent [If Q.9 answer is \quotes{I received a form of alert or notification from my bank or card issuer...} or \quotes{I received a form of alert or notification from a third-party...}] \\ 
\noindent \textbf{Q.11:} Through what means did you receive the notification? (select all that apply)
\begin{itemize}[noitemsep,topsep=0pt]
    \item Email
    \item Post letter
    \item Phone call
    \item Automated voice message
    \item Short text message (SMS)
    \item Message through the bank or card issuer’s  app or Internet banking website
    \item Push notification through the bank or card issuer’s app or Internet banking website
    \item Other (please specify): [free response field]
\end{itemize}

\noindent [If Q.10 answer is \quotes{Suspicious transaction} or \quotes{Fraudulent transaction}] \\
\noindent \textbf{Q.12:} Did the bank or card issuer take any action to stop the fraud automatically?
\begin{itemize}[noitemsep,topsep=0pt]
    \item Yes
    \item No
    \item I cannot remember
\end{itemize}

\noindent [If Q.9 answer is \quotes{I received a form of alert or notification from my bank or card issuer...} or \quotes{I received a form of alert or notification from a third-party...}]  \\
\noindent \textbf{Q.13:} \ul{\textbf{From the time your debit/credit card fraudulent transaction occurred}}, how long did it take \textbf{the notification to reach you}? \\
\noindent \textit{If the fraud incident consisted of multiple fraudulent transactions, please answer based on the first transaction.}
\begin{itemize}[noitemsep,topsep=0pt]
    \item Immediately
    \item Within a few minutes
    \item Within an hour
    \item Within a few hours
    \item Within a day
    \item Within a few days
    \item Within a week
    \item Within a few weeks
    \item Within a month
    \item More than a month
    \item I cannot remember 
    \item Other (please specify): [free response field]
\end{itemize}

\noindent [If Q.9 answer is \quotes{I received a form of alert or notification from my bank or card issuer...} or \quotes{I received a form of alert or notification from a third-party...}]  \\
\noindent \textbf{Q.14:} Overall, how helpful or unhelpful were the notifications you received in detecting the fraudulent activity?
\begin{itemize}[noitemsep,topsep=0pt]
    \item Extremely helpful
    \item Somewhat helpful
    \item Slightly helpful
    \item Not at all helpful 
\end{itemize}

\noindent [If Q.9 answer is \quotes{I received a form of alert or notification from my bank or card issuer...} or \quotes{I received a form of alert or notification from a third-party...}]  \\
\noindent \textbf{Q.15:} Why did you think the notification was helpful or unhelpful?\\  \noindent [Free response field]


\noindent \textbf{Q.16:} \ul{\textbf{From the time your debit/credit card fraudulent transaction occurred}}, how long did it take \textbf{you to realize that you had been defrauded}? \\ 
\noindent \textit{If the fraud incident consisted of multiple fraudulent transactions, please answer based on the first transaction}
\begin{itemize}[noitemsep,topsep=0pt]
    \item Immediately
    \item Within a few minutes
    \item Within an hour
    \item Within a few hours
    \item Within a day
    \item Within a few days
    \item Within a week
    \item Within a few weeks
    \item Within a month
    \item More than a month
    \item I cannot remember 
    \item Other (please specify): [free response field]
\end{itemize}

\noindent \textbf{Q.17:} How many fraudulent transactions occurred in that incident before you were able to detect it?
\begin{itemize}[noitemsep,topsep=0pt]
    \item One transaction 
    \item Multiple transactions in the same day
    \item Multiple transactions over a period of time 
    \item I cannot remember 
    \item Other (please specify): [free response field]
\end{itemize}

\noindent \textbf{Q.18:} Do you have any suggestions for how your bank or card issuer could improve fraud alerting and detection capabilities?\\ 
\noindent [Free response field]~\\

\noindent SECTION \#3: REPORTING

\noindent Please think only about the time of \textbf{your most recent experience of debit/credit card fraud that you described previously} in this survey, and answer all the following questions in that context.

\noindent \textbf{Q.19:} Did you report the fraud incident to your debit/credit card’s bank or card issuer?
\begin{itemize}[noitemsep,topsep=0pt]
    \item Yes
    \item No
    \item I cannot remember
\end{itemize}

\noindent [If Q.19 answer is \quotes{no}] \\
\noindent \textbf{Q.20:} Why did you not report this fraud incident to the bank or card issuer? \\ 
\noindent [Free response field]

\noindent [If Q.19 answer is \quotes{yes}] \\
\noindent \textbf{Q.21:} 
Did you seek compensation for the amount the fraudster charged to your card (the direct financial loss) from your bank or card issuer? 
\begin{itemize}[noitemsep,topsep=0pt]
    \item Yes
    \item No
    \item I cannot remember
\end{itemize}

\noindent [If Q.19 answer is \quotes{yes}] \\
\noindent \textbf{Q.22:} Have you been compensated for the amount the fraudster charged to your card (the direct financial loss) by your bank or card issuer?
\begin{itemize}[noitemsep,topsep=0pt]
    \item Yes - Fully
    \item Yes - Partially
    \item No
    \item Other (please specify): [free response field]
\end{itemize}

\noindent [If Q.19 answer is \quotes{yes}] \\
\noindent \textbf{Q.23:} Did the bank or card issuer provide you with an explanation of the fraud incident \textbf{(what / how / where / by whom, the incident was)}?
\begin{itemize}[noitemsep,topsep=0pt]
    \item The bank or card issuer provided me with \textbf{full explanation} of the fraud details
    \item The bank or card issuer provided me with \textbf{partial explanation} of the fraud details
    \item The bank or card issuer \textbf{did not provide} me with \textbf{any explanation} of the fraud details
    \item Other (please specify): [free response field]
\end{itemize}

\noindent [If Q.19 answer is \quotes{yes}] \\
\noindent \textbf{Q.24:} How satisfied or unsatisfied were you with the support offered to you by your bank or card issuer in handling the incident’s reporting and compensation processes?
\begin{itemize}[noitemsep,topsep=0pt]
    \item Extremely satisfied
    \item Somewhat satisfied
    \item Slightly satisfied
    \item Not at all satisfied 
\end{itemize}

\noindent [If Q.19 answer is \quotes{yes}] \\
\noindent \textbf{Q.25:} What advice would you give your bank or card issuer to improve the support they offered to you in handling the incident’s reporting and compensation processes?\\ 
\noindent [Free response field]~\\

\noindent SECTION \#4: FRAUD IMPACT \\
\noindent Please think only about the time of \textbf{your most recent experience of debit/credit card fraud that you described previously} in this survey, and answer all the following questions in that context.

\noindent \textbf{Q.26:} How much did \textbf{the fraudster charge to your card} (the direct financial loss)? \ul{\textbf{Please select the total amount, even if it was later compensated or reimbursed}}. 
\begin{itemize}[noitemsep,topsep=0pt]
    \item \$0 (no direct financial loss)
    \item From \$0.01 to \$10
    \item From \$11 to \$50
    \item From \$51 to \$100
    \item From \$101 to \$500
    \item From \$501 to \$1,000
    \item From \$1,001 to \$2,000
    \item From \$2,001 to \$4,000
    \item From \$4,001 to \$6,000
    \item From \$6,001 to \$8,000
    \item From \$8,001 to \$10,000
    \item More than \$10,000
    \item I cannot remember
    \item Other (please specify): [free response field]
\end{itemize}

\noindent \textbf{Q.27:} Which, if any, of the following indirect financial losses did you experience as a result of the fraud experience you had? (select all that apply)
\begin{itemize}[noitemsep,topsep=0pt]
    \item Damaged personal reputation 
    \item Loss of trust from others
    \item Legal and attorney fees for pursuing legal action or defending against accusations
    \item Increased insurance premiums or difficulty obtaining insurance coverage
    \item Costs of credit monitoring or identity theft protection services
    \item Loss of job due to the fraud incident
    \item Loss of income due to the fraud incident
    \item Emotional distress
    \item Costs of therapy or counseling
    \item Difficulty obtaining loans or credit
    \item Negative impact on credit score or higher interest rates for borrowing
    \item None of these [exclusive]
    \item Other (please specify): [free response field]
\end{itemize}

\noindent \textbf{Q.28:} Which, if any, of the following psychological negative impacts did you experience as a result of the fraud experience you had? (select all that apply)
\begin{itemize}[noitemsep,topsep=0pt]
    \item Stress
    \item Difficulty in sleeping
    \item Difficulty in trusting others
    \item Confusion
    \item Discomfort
    \item Frustration
    \item Worry or anxiety
    \item Feeling upset
    \item Feeling unsafe
    \item Depression
    \item Embarrassment
    \item Shame
    \item Guilt
    \item Loss of self-confidence
    \item Low satisfaction
    \item Negative changes in perception
    \item None of these [exclusive]
    \item Other (please specify): [free response field]
\end{itemize}

\noindent \textbf{Q.29:} How did the fraud incident impact you \textbf{at the time of the incident}? \\ 
\noindent [\autoref{tab:matrix_attime} is shown here.]
\begin{table*}[h!]
    \centering
    \caption{Impact matrix used in questions Q29 and Q30.}
    \label{tab:matrix_attime}
    \resizebox{0.70\textwidth}{!}{
    \begin{tabular}{l|c|c|c}
    \toprule
                        &  \makecell[l]{Did not impacted \\me negatively at all} & \makecell[l]{Somewhat impacted \\me negatively} & \makecell[l]{Strongly impacted me \\negatively} \\
    \midrule
        Financially     &   $\circ$       &    $\circ$       &    $\circ$       \\
        Psychologically  &  $\circ$        &   $\circ$       &   $\circ$        \\
        \makecell[l]{Your level of trust in \\performing financial transactions} &   $\circ$       &  $\circ$         &  $\circ$         \\
    \bottomrule    
    \end{tabular}
  } 
\end{table*}

\noindent \textbf{Q.30:} How does the fraud incident continue to impact you \textbf{today}? \\
\noindent [\autoref{tab:matrix_attime} is shown here.]~\\

\noindent SECTION \#5: BEHAVIORAL CHANGE \\
\noindent Please think only about the time of \textbf{your most recent experience of debit/credit card fraud that you described previously }in this survey, and answer all the following questions in that context.

\noindent \textbf{Q.31:} Have you taken any measures to prevent future fraud on your debit/credit card? (select all that apply) 
\begin{itemize}[noitemsep,topsep=0pt]
    \item I have implemented security measures from the bank or card issuer to protect my financial transactions.
    \item I have implemented security measures from third-parties to protect my financial transactions. 
    \item I am now more cautious and vigilant when conducting financial transactions.
    \item I have set up alerting mechanisms for my account or card.
    \item I now regularly check my account or card statement.
    \item I switched my bank or card issuer.
    \item I did not undertake any measures [exclusive]
    \item Other (please specify): [free response field]
\end{itemize}

\noindent [If Q.31 answer is NOT “I did not undertake any measures"] \\
\noindent \textbf{Q.32:} Please elaborate more on the measures you took to prevent future fraud on your debit/credit card? \\
For your reference, you said you took the following measure(s) in the previous answer.\\
\noindent [Response to Q.31 is shown as bullet list] \\
\noindent [Free response field]~\\

\noindent SECTION \#6: DEMOGRAPHICS \newline 
\noindent \textbf{Please answer the following demographic questions}

\noindent \textbf{Q.33:} How do you describe your gender identity?
\begin{itemize}[noitemsep,topsep=0pt]
    \item Female
    \item Male 
    \item Non-binary
    \item Prefer to self-describe: [free response field]  
    \item Prefer not to answer
\end{itemize}

\noindent \textbf{Q.34:} How old are you (in years)?
\begin{itemize}[noitemsep,topsep=0pt]
    \item From 18 to 24 
    \item From 25 to 34
    \item From 35 to 44
    \item From 45 to 54
    \item From 55 to 64
    \item From 65 to 74
    \item 75 or older
    \item Prefer not to answer
\end{itemize}

\noindent \textbf{Q.35:} What is your race or ethnic identity? (select all that apply)
\begin{itemize}[noitemsep,topsep=0pt]
    \item White
    \item Black or African American
    \item American Indian or Alaska Native
    \item Asian
    \item Native Hawaiian or Pacific Islander
    \item Hispanic and/or Latino/Latina/Latinx
    \item Prefer not to answer [exclusive]
    \item Other (please specify): [free response field]
\end{itemize}

\noindent \textbf{Q.36:} What is the highest educational degree you have received?
\begin{itemize}[noitemsep,topsep=0pt]
    \item Doctoral degree
    \item Master’s degree
    \item Bachelor’s degree
    \item Associate’s degree
    \item High school diploma or GED
    \item Less than high school degree
    \item Other (please specify): [free response field]
\end{itemize}

\noindent \textbf{Q.37:} Do you have a university degree in, or currently work in, one or more of the following fields: Computer Science (CS), Information Systems (IS), Information Technology (IT), or Computer Engineering (CE)?
\begin{itemize}[noitemsep,topsep=0pt]
    \item Yes
    \item No
\end{itemize}

\noindent \textbf{Q.38:} What is your current employment status?
\begin{itemize}[noitemsep,topsep=0pt]
    \item Student
    \item Full-time employee
    \item Part-time employee
    \item Self-employed or business owner
    \item Full-time homemaker
    \item Unemployed, and looking for a job
    \item Unemployed, and not looking for a job
    \item Unable to work
    \item Retired
    \item Other (please specify): [free response field]
\end{itemize}

\noindent \textbf{Q.39:} What is your approximate annual household income? \\
\noindent \textit{Please answer based on your entire current household's income, before taxes.}
\begin{itemize}[noitemsep,topsep=0pt]
    \item Less than \$20,000
    \item \$20,000 to \$39,999
    \item \$40,000 to \$59,999
    \item \$60,000 to \$79,999
    \item \$80,000 to \$99,999
    \item \$100,000 to \$149,999
    \item \$150,000 or more
    \item Prefer not to answer
\end{itemize}

\noindent [If Q.38 answer is \quotes{full-time employee} or \quotes{part-time employee} or \quotes{Self-employed or business owner}]\\
\noindent \textbf{Q.40:} What is the sector you currently work in?
\begin{itemize}[noitemsep,topsep=0pt]
    \item Pre-university Education
    \item University Education
    \item Health
    \item Communication and Information Technology
    \item Financial
    \item Industrial
    \item Agricultural
    \item Sales and retail
    \item Petrochemical
    \item Other (please specify): [free response field]
\end{itemize}

\noindent [If Q.38 answer is \quotes{full-time employee} or \quotes{part-time employee} or \quotes{Self-employed or business owner}]\\
\noindent \textbf{Q.41:} What is your current job title? (e.g. Teacher, Assistant Professor, Administrative staff, Nurse, etc.) \\
\noindent [Free response field]

\noindent [If Q.38 answer is \quotes{student}]\\
\noindent \textbf{Q.42:} What is the degree you are currently studying?
\begin{itemize}[noitemsep,topsep=0pt]
    \item Doctoral degree
    \item Master’s degree
    \item Bachelor’s degree
    \item Associate’s degree
    \item High school diploma or GED
    \item Less than high school degree
    \item Other (please specify): [free response field]
\end{itemize}

\noindent [If Q.38 answer is \quotes{student}]\\
\noindent \textbf{Q.43:} What is the major you are studying for your current degree? \\
\noindent [Free response field]

\noindent \textbf{Q.44:} Do you have a university degree in cybersecurity or currently work in the cybersecurity area?
\begin{itemize}[noitemsep,topsep=0pt]
    \item Yes
    \item No
\end{itemize}

\noindent \textbf{Q.45:} If you have any other thoughts or feedback about this survey or the information you viewed, please let us know here. (optional) \\
\noindent [Free response field]

\noindent \textbf{You now reached the end of the survey. To submit your response click the \quotes{Submit} button}.

\clearpage
\onecolumn
\section{Codebook} \label{app:codebook}
In this section, we list our codebook divided into tables. Each table represents the codebook for a question or a set of related questions. All question are from the main survey listed in~\autorefappendix{app:main_survey} unless stated otherwise. All codes are in small letters only.

\begin{table*}[!h]
\centering
\footnotesize
\caption{Codebook for Q.6 of the screening survey about fraud description (transaction mode).}
\label{tab:preferred}
\renewcommand{\arraystretch}{1.25}
\resizebox{\textwidth}{!}{%
    \begin{tabular}{p{0.3\textwidth}|p{0.7\textwidth}}
    \toprule 
    \rowcolor{gray!50}
    \multicolumn{2}{l}{Q.6 (screening survey): Fraud description $\rightarrow$ Transaction mode} \\
    \midrule
    \textbf{Code (\# occurrence) [exclusive]}   & \textbf{Code Definition}  \\
    \hline
    unspecified (78)         & No cues about the transaction type \\  
    \hline 
    online (53)              & Cues indicating an online transaction \\
    \hline 
    offline (19)             & Cues indicating an offline transaction  \\ 
    \bottomrule
    \end{tabular}
} 
\end{table*}

\begin{table*}[!h]
\centering
\footnotesize
\caption{Codebook for Q.6 of the screening survey about fraud description (transaction category).}
\label{tab:template}
\renewcommand{\arraystretch}{1.25}
\resizebox{\textwidth}{!}{%
    \begin{tabular}{p{0.3\textwidth}|p{0.7\textwidth}}
    \toprule 
    \rowcolor{gray!50}
    \multicolumn{2}{l}{Q.6 (screening survey): Fraud description $\rightarrow$ Transaction category} \\
    \midrule
    \textbf{Code (\# occurrence) [non-exclusive]}   & \textbf{Code Definition} \\ 
    \hline
    unauthorized\_purchase (87)   & Transactions where any goods were purchased from any store in a any location \\
    \hline 
    unauthorized\_transaction (54)        & A generic code for transactions/charges made to a bank account or cards, where there isn't a specific indicator or type mentioned \\
    \hline 
    overcharging (4)       & Transactions including purchases, tips, etc. which charged more than the authorized amount  \\
    \hline 
    identity\_theft (3)       & Transactions where the participant's bank account information was stolen or a new bank account or credit card was opened in their name without their knowledge or consent \\
    \hline 
    unauthorized\_withdrawal (2)       & Transactions which explicitly mentioned withdrawal of amount from their account  \\
    \hline 
    never\_received\_items (1)       & Transactions where a purchase was made but the participant never received the purchased item  \\
    \hline 
    unauthorized\_check\_issued (1)       & Frauds associated with cash access checks \\
    \bottomrule
    \end{tabular}
} 
\end{table*}

\begin{table*}[!h]
\centering
\footnotesize
\caption{Codebook for Q.6 of the screening survey about fraud description (location).}
\label{tab:template}
\renewcommand{\arraystretch}{1.25}
\resizebox{\textwidth}{!}{%
    \begin{tabular}{p{0.3\textwidth}|p{0.7\textwidth}}
    \toprule 
    \rowcolor{gray!50}
    \multicolumn{2}{l}{Q.6 (screening survey): Fraud description $\rightarrow$ Transaction location} \\
    \midrule
    \textbf{Code (\# occurrence) [exclusive]}   & \textbf{Code Definition}   \\
    \hline
    unspecified (116)        & Transaction that did not specify location \\
    \hline 
    different\_state (23)         & Transaction occurred in a state different than the participants' state (within USA) \\
    \hline 
    different\_continent (4)       & Transaction occurred outside of North America  \\
    \hline 
    different\_area (3)         &  Transaction occurred in an area different than the participants' area (within USA)  \\
    \hline 
    different\_city (2)         &  Transaction occurred in a city different than the participants' city (within USA) \\
    \hline 
    different\_country (2)         & Transaction occurred in a country different than the participants' country (USA)  \\
    \bottomrule
    \end{tabular}
} 
\end{table*}

\begin{table*}[!h]
\centering
\footnotesize
\caption{Codebook for Q.6 of the screening survey about fraud description (goods or service).}
\label{tab:template}
\renewcommand{\arraystretch}{1.25}
\resizebox{\textwidth}{!}{%
    \begin{tabular}{p{0.3\textwidth}|p{0.7\textwidth}}
    \toprule 
    \rowcolor{gray!50}
    \multicolumn{2}{l}{Q.6 (screening survey): Fraud description $\rightarrow$ Goods or service} \\
    \midrule
    \textbf{Code (\# occurrence) [non-exclusive]}   & \textbf{Code Definition}  \\
    \hline
    store (22)              & Transaction that specified a store \\
    \hline 
    food\_and\_beverages (19)       & Transaction about purchasing food or beverages \\
    \hline 
    electronics (7)        &  Transaction about purchasing electronics or services supporting these electronics (e.g. online gaming account) or stores that primarily sell electronics  \\
    \hline 
    gas (7)                 &  Transaction about purchasing  gas or occurred in a gas station  \\
    \hline 
    subscription\_membership (6)     & Transaction about purchasing/renewing a subscription or membership \\
    \hline 
    clothes (4)            & Transaction about purchasing clothes in an online/offline mode \\
    \hline 
    entertainment (3)      & Transaction about purchasing access to content primarily for entertainment including online avenues such as Amazon Prime Video or offline avenues such as wrestling or record companies \\
    \hline 
    transportation (3)     &  Transaction about purchasing tickets for transportation or ride-sharing app transactions  \\
    \hline 
    horse\_equipment (1)    & Transaction about purchasing equipment for horses \\
    \hline 
    like\_money\_laundering\_website (2)   &  Transaction about money laundering  \\
    \hline 
    loan\_agreement (1)    & Transaction about loan agreement \\
    \hline 
    porn (1)                & Transaction about porn  \\
    \bottomrule
    \end{tabular}
} 
\end{table*}

\begin{table*}[!h]
\centering
\footnotesize
\caption{Codebook for Q.6 of the screening survey about fraud description (special cases).}
\label{tab:template}
\renewcommand{\arraystretch}{1.25}
\resizebox{\textwidth}{!}{%
    \begin{tabular}{p{0.3\textwidth}|p{0.7\textwidth}}
    \toprule 
    \rowcolor{gray!50}
    \multicolumn{2}{l}{Q.6 (screening survey): Fraud description $\rightarrow$ Special cases} \\
    \midrule
    \textbf{Code (\# occurrence) [non-exclusive]}   & \textbf{Code Definition}  \\
    \hline
    unspecified (115)                   & Transaction that did not specify special cases \\ 
    \hline 
    online\_account (14)         & Online account hacked to make purchases they did not make \\
    \hline 
    card\_details (8)           & Card details were accessed in an unauthorized manner  \\
    \hline 
    service\_provider\_notification (7)        &  When a notification originated from the service provider (e.g. store) \\
    \hline 
    physical\_card (6)         & Someone took their physical card or smartphone \\
    \hline 
    smartphone (1)             & Physical theft for a smartphone that contains the app that contains the card  \\
    \hline 
    family (1)                 & Family was involved in the fraud \\
    \hline 
    skimming (1)              &  Card skimming was involved in the fraud  \\
    \bottomrule
    \end{tabular}
} 
\end{table*}

\begin{table*}[!h]
\centering
\footnotesize
\caption{Codebook for Q.15 about reasons for notification helpfulness.}
\label{tab:template}
\renewcommand{\arraystretch}{1.25}
\resizebox{\textwidth}{!}{%
    \begin{tabular}{p{0.3\textwidth}|p{0.7\textwidth}}
    \toprule 
    \rowcolor{gray!50}
    \multicolumn{2}{l}{Q.\#: Notifications helpfulness / unhelpfulness $\rightarrow$ Reasons for notification helpfulness } \\
    \midrule
    \textbf{Code (\# occurrence) [non-exclusive]}   & \textbf{Code Definition}   \\
    \hline
    awareness (49) & The notification made the subject aware of the fraud, such as being alerted, notified, etc. \\
    \hline
    fast (29) &  The notification enabled fast awareness, action, or stopping of the fraud, such as within \# [time unit], immediately, quickly, etc. \\
    \hline
    helped\_take\_action (29) & The notification enabled the subject to take actions based on on the notification to stop the current or future fraud \\
    \hline
    transaction\_type (5) &  The notification contained the transaction type (e.g. purchase) and number of transactions (single, multiple, etc.) \\
    \hline
    card\_used (2) & The notification alerted the subject that their card was used \\
    \hline
    info\_compromise (2) &  The notification alerted the subject that their information was compromised or compromised in a data breach \\
    \hline
    late (2) &  The notification arrived late \\
    \hline
    place (2) &  The notification provided the place where the card used \\
    \hline
    suspicious\_transaction (2) &  The notification alerted the subject about suspicious transaction explicitly \\
    \hline
    amount (1) & The notification provided the amount taken \\
    \hline
    confidence\_in\_service (1) &  The notification allowed them gain more confidence on the service \\
    \hline
    fear\_phishing\_msgs (1) &  The subject mentioned fear of phishing attacks in notifications \\
    \hline
    helped\_monitoring (1) &  The notification helped monitoring accounts/cards \\
    \hline
    helped\_resolved (1) & The notification helped in resolving the fraud\\
    \hline
    more\_info (1) & The notification provided subsequent information about the incident (more than notification basic information)\\
    \hline
    no\_loss\_access (1) &  The subject mentioned they did not lose access to their accounts/cards\\
    \hline
    stopped\_psych\_harm (1) & The notification helped protect the subject from psychological harm such as embarrassment\\
    \bottomrule
    \end{tabular}
} 
\end{table*}

\begin{table*}[!h]
\centering
\footnotesize
\caption{Codebook for Q.15 of the main survey about reasons for notification unhelpfulness.}
\label{tab:template}
\renewcommand{\arraystretch}{1.25}
\resizebox{\textwidth}{!}{%
    \begin{tabular}{p{0.3\textwidth}|p{0.7\textwidth}}
    \toprule 
    \rowcolor{gray!50}
    \multicolumn{2}{l}{Q.\#: Notifications helpfulness / unhelpfulness $\rightarrow$ Reasons for notification unhelpfulness } \\
    \midrule
    \textbf{Code (\# occurrence) [non-exclusive]}   & \textbf{Code Definition}   \\
    \hline
    fraud\_occurred (1) & The fraud had already occurred \\
    \bottomrule
    \end{tabular}
} 
\end{table*}

\begin{table*}[!h]
\centering
\footnotesize
\caption{Codebook for Q.18 about suggestions to improve alert and detection capabilities.}
\label{tab:template}
\renewcommand{\arraystretch}{1.25}
\resizebox{\textwidth}{!}{%
    \begin{tabular}{p{0.3\textwidth}|p{0.7\textwidth}}
    \toprule 
    \rowcolor{gray!50}
    \multicolumn{2}{l}{Q.18: Detection $\rightarrow$ Suggestion to improve alert and detection capabilities} \\
    \midrule
    \textbf{Code (\# occurrence) [non-exclusive]}   & \textbf{Code Definition}     \\
    \hline
    different\_location\_alert (16) & 	Send alerts of transactions initiated from geographical locations different than the card holder's resident location \\
    \hline 
    sms\_alert (15) & Send alerts through SMS \\
    \hline 
    approve\_decline\_transaction (12)	& Allow individuals to approve or decline certain transactions \\
    \hline 
    fraud\_or\_suspicious\_transaction\_alert (11)  & Send alerts of suspicious or fraudulent transactions \\
    \hline 
    immediate (11) &  Send alerts immediately \\ 
    \hline 
    unusual\_behavior\_alert (11) & Send alerts of unusual behaviour \\
    \hline 
    purchase\_alert (8) &	Send alerts for every purchase transaction \\
    \hline
    email\_alert (7) &	Send alerts through email \\
    \hline  
    multi\_channel\_alerts (6) &	Send alerts through multiple channels (e.g. phone/sms/email/etc.) \\
    \hline 
    large\_purchase\_alert (6) &	Send alerts for large transactions \\
    \hline 
    any\_transaction\_alert (5) &	Send alerts for any type of transaction \\
    \hline 
    phone\_call\_alert (5) &	Send alerts through phone call \\
    \hline 
    app\_notification (3) &		Send alerts through app notifications \\
    \hline 
    int\_transaction\_alert (2) &	Send alerts for international or overseas transactions \\
    \hline 
    alert\_without\_cardlock (2) &	Send alerts instead of locking the card \\
    \hline 
    default\_alert (2) &	Send alerts by default without having to opt-in \\
    \hline 
    legitimate\_text (2) &		Ensure the alert themselves do not look fraudulent or suspicious \\
    \hline 
    multi\_channel\_verification (2) &	Verify transaction through multiple channels (e.g. phone/call/email) \\
    \hline 
    phone\_alert (2)	        & Send alerts on the phone device \\
    \hline 
    require\_verification (2) &		Require verification of transaction \\
    \hline 
    additional\_info (1) &		Provide additional information in alerts \\
    \hline 
    blacklist\_fraudulent\_sellers (1) &  Maintain a blacklist of fraudulent sellers and block transaction toward them \\
    \hline 
    customizable\_alerts (1) &		Allow customization of alerts  \\
    \hline 
    enforce\_pin\_online (1) &		Require pin for purchases online \\
    \hline 
    enforce\_pin (1) 	   & Require pin for purchases offline \\
    \hline 
    giftcard\_transaction\_alert (1) &		Send alerts for gift card purchasing transactions \\
    \hline 
    help\_find\_offenders (1) &		Find the the fraudsters who caused the fraud \\
    \hline 
    help\_prosecute\_offenders (1) &	Prosecute the fraudsters who caused the fraud \\
    \hline 
    improve\_online\_purchasing\_security (1)  &	Improve security features/functionalities/measures offered by the bank or card issuer for online purchases \\
    \hline 
    improve\_security (1) &	Improve security features/functionalities/measures offered by the bank or card issuer \\
    \hline 
    lock\_card\_feature (1) &	Offer consumers the ability to lock/unlock card by themselves \\
    \hline 
    lock\_upon\_no\_response\_to\_alert (1) & Lock the card if there isn't a response from the card holder after a predefined amount of time \\
    \hline 
    monitor\_small\_transactions (1) &	Monitor transactions that involve small amounts \\
    \hline 
    online\_reporting (1) & Provide means to report the fraud online \\
    \hline 
    optional\_alert (2)	 & Send alerts if the card holder opted for one \\
    \hline 
    password\_change\_alert (1) &	Send alerts if the bank account or cards password has changed \\
    \hline 
    pin\_change\_alert (1)	& Send alerts if the bank account or cards pin number has changed \\
    \hline 
    reduce\_false\_positives (2) &	Reduce alerts for non-fraudulent transactions (flase positive)\\
    \hline 
    remind\_to\_configure (1) &	 Remind users to configure alerts/notifications \\
    \hline 
    repeated\_purchases\_alert (1) &	Send alerts upon repeated purchase of the same goods or services \\
    \hline 
    skimming\_detection\_alert (1) &	Send alerts if the users card was subject to card skimming \\
    \hline 
    stop\_mailing\_cash\_access\_checks (1) &	Stop mailing cash access checks without explicit request or authorization  \\
    \hline 
    track\_card (1) &		Track the location of the card \\
    \hline 
    track\_transaction\_recipients (1) & Track the recipients of the transaction \\
    \hline 
    over\_limit\_transaction\_alert (1) &	Send alerts for transactions over a specified limit \\
    \bottomrule
    \end{tabular}
} 
\end{table*}

\begin{table*}[!h]
\centering
\footnotesize
\caption{Codebook for Q.25 about suggestion to improve the reporting and compensation process.}
\label{tab:template}
\renewcommand{\arraystretch}{1.25}
\resizebox{\textwidth}{!}{%
    \begin{tabular}{p{0.3\textwidth}|p{0.7\textwidth}}
    \toprule 
    \rowcolor{gray!50}
    \multicolumn{2}{l}{Q.25: Reporting $\rightarrow$ Suggestion to improve the reporting and compensation process.} \\
    \midrule
    \textbf{Code (\# occurrence) [non-exclusive]}   & \textbf{Code Definition}     \\
    \hline
    fraud\_explanation (26) & The bank or card issuer provides an explanation of the fraud, such as how the fraud occurred, who the recipient/fraudster was, when and where did the incident occur, etc. \\
    \hline 
    refund (14) &		The bank or card issuer provides refund to the victims\\
    \hline 
    fast\_process (9) &		The bank or card issuer provides faster process to report and resolve the fraud \\
    \hline 
    fast\_refund (8) &		The bank or card issuer provides faster refund to the victims \\
    \hline 
    fast\_comm (7)	&	The bank or card issuer provides faster communication to the victims \\
    \hline 
    full\_refund (4) &		The bank or card issuer provides complete/full refund of the amount fraudulently charged \\
    \hline 
    preventive\_tips (5)	&	The bank or card issuer provides tips to prevent such fraud in the future \\
    \hline 
    automated\_reporting\_system (3) &		The bank or card issuer provides an automated process to report and resolve the fraud \\
    \hline 
    better\_comm (3) &		The bank or card issuer communicates with the affected parties in a better manner \\
    \hline 
     proactive (3) &		The bank or card issuer takes proactive action in identifying fraudulent transactions and resolving them \\
    \hline 
     trust\_customer (3) &		The bank or card issuer trusts the consumer \\
    \hline 
    direct\_comm (2) &		The bank or card issuer allows direct communication with a human representative \\
    \hline 
    resolution (2) &	The bank or card issuer provides a resolution to the fraud experienced by the victims \\
    \hline 
    approve\_decline\_transaction (2) &	 The bank or card issuer provides means to approve or decline transactions \\
    \hline 
    easy\_freeze\_card (2) &	The bank or card issuer provides means to freeze the victim's card \\
    \hline 
    fast\_alert (2)	&	The bank or card issuer alerts the victims in a faster manner\\
    \hline 
    fraud\_or\_suspicious\_transaction\_alert (2) & The bank or card issuer alerts the victims in the event of fraudulent or suspicious transaction \\
    \hline 
    understanding (1) &		Representatives from the bank or card issuer showcase understanding with the consumer in the incident reporting/resolution process \\
    \hline 
    additional\_compensation (1) &		The bank or card issuer provides additional compensation in addition to the reimbursement of the fraudulently acquired amount \\
    \hline 
    avoid\_mistakes (1) &		The bank or card issuer avoids making mistakes in the reporting/resolution process \\
    \hline 
    better\_emp\_training (1) &		The bank or card issuer provides better training to their employees \\
    \hline 
    better\_emp\_wages (1) &		The bank or card issuer provides better compensation to their employees \\
    \hline 
    black\_list\_fraud\_recipient (1) &		The bank or card issuer maintains a blacklist of fraudulent sellers and block transactions toward them \\
    \hline 
    fast\_new\_card (1) &		The bank or card issuer provides a replacement card in a faster manner \\
    \hline 
    fast\_resolution (1) &		The bank or card issuer provides faster resolution to the victims \\
    \hline 
    fraud\_detection (1) &		The bank or card issuer provides better fraud detection capabilities \\
    \hline 
    fraud\_explanation\_from\_retailer (1) &		The retailers/third-party service provider provides an explanation of the fraud, such as how the fraud occurred, who the recipient/fraudster was, when and where did the incident occur, etc. \\
    \hline 
    keep\_same\_card (1) &		The bank or card issuer retains the same card, and not have to get a new card \\
    \hline 
    multi\_channel\_comm (1) &		The bank or card issuer contacts the victims through multiple channels \\
    \hline 
    patience (1) &	Representatives from the bank or card issuer showcase patience with the consumer in the incident reporting/resolution process \\
    \hline 
    prosecute\_fraudster (1)	& The bank or card issuer helps prosecute the fraudster  \\
    \hline 
    reporting\_via\_app (1) &		The bank or card issuer provides means to report fraud through the bank or card issuer's mobile app \\
    \hline 
    reporting\_via\_website (1) &	The bank or card issuer provides means to report fraud through the bank or card issuer's website \\
    \hline 
    use\_email (1) &		The bank or card issuer contacts victims or share alerts through email \\
    \hline 
    use\_mfa (1) &		The bank or card issuer provides authentication checks with multi-factor authentication \\
    \hline 
    use\_sms (1) &		The bank or card issuer provides authentication checks via SMS \\
    \bottomrule
    \end{tabular}
} 
\end{table*}
\begin{table*}[!h]
\centering
\footnotesize
\caption{Codebook for Q.25 about suggestion to improve the reporting and compensation process (factors that led to satisfaction).}
\label{tab:template}
\renewcommand{\arraystretch}{1.25}
\resizebox{\textwidth}{!}{%
    \begin{tabular}{p{0.3\textwidth}|p{0.7\textwidth}}
    \toprule 
    \rowcolor{gray!50}
    \multicolumn{2}{l}{Q.25: Reporting $\rightarrow$ Factors that led to satisfaction.} \\
    \midrule
    \textbf{Code (\# occurrence) [non-exclusive]}   & \textbf{Code Definition}     \\
    \hline
   satisfied (17) & The participant was satisfied with the reporting and resolution process \\
   \hline 
   fast\_process(10) & Fast reporting and resolution process \\
   \hline 
   refund (5) & Refund was provided\\
    \hline
    comm (2) & Communication efforts by the bank or card issuer representative \\
    \hline 
    full\_refund (2) & Full refund was provided\\
    \hline
   provided\_resolution (1) & Resolution was provided\\
   \hline 
    professionalism (1) & Professionalism exhibited by the bank or card issuer representative \\
    \hline 
    fast\_comm (1) & Fast communication by the bank or card issuer \\
    \hline 
    provided\_direct\_comm (1) & Direct communication efforts by the bank or card issuer human representative \\
    \hline 
    fast\_refund (1) & Refund was provided in a fast manner \\
    \hline 
    new\_card (1) & New card was provided \\
    \hline 
    provided\_support (1) & Support was provided by the bank or card issuer \\
    \hline 
    fraud\_prevented (1) & The fraud was prevented by the bank or card provider \\
    \bottomrule
    \end{tabular}
} 
\end{table*}

\begin{table*}[!h]
\centering
\footnotesize
\caption{Codebook for Q.25 about suggestion to improve the reporting and compensation process (fraud explanation).}
\label{tab:template}
\renewcommand{\arraystretch}{1.25}
\resizebox{\textwidth}{!}{%
    \begin{tabular}{p{0.3\textwidth}|p{0.7\textwidth}}
    \toprule 
    \rowcolor{gray!50}
    \multicolumn{2}{l}{Q.25: Reporting $\rightarrow$ Fraud explanation.} \\
    \midrule
    \textbf{Code (\# occurrence) [exclusive]}   & \textbf{Code Definition}     \\
    \hline
    culprit\_info (7) &  More information regarding who the culprit was \\
    \hline 
    incident\_details (5) & More information regarding the details of transaction that caused the incident \\
    \hline 
    what\_happened (5) & More information regarding what happened in this incident \\
    \hline 
    how\_happened (4) &  More information regarding the how the incident occurred \\
    \hline 
    preventive\_tips (4) & More information on how to prevent such incident from happening in the future \\
    \hline 
    investigation\_details (2) & More information regarding the details of investigation conducted \\
    \hline 
    legal\_action (2) &  Aid in legal action undertaken by the affected party \\
    \hline 
    where\_happened (2) & More information regarding the where the incident occurred \\
    \hline 
    assessment (1) & Conduct an accurate assessment of the situation \\
    \hline 
    automate\_reporting (1) &	Automated means to report the incident \\
    \hline
    online\_or\_offline (1) &	More information regarding whether the transaction was online or offline \\
    \hline 
    process\_info (1) &		More information regarding the process of incident reporting and resolution \\
    \hline
    recipient\_info (1) & More information regarding who the recipient of this transaction/amount was \\
    \hline 
    report\_app\_website (1) & Means to report the incident through the mobile app or website \\
    \hline 
    support\_channels (1) & More information regarding the support channels available for reporting the incident and seeking resolution \\
    \hline 
    when\_happened (1) &	 More information regarding the when the incident occurred \\ 
    \hline 
    why\_happened (1) &		More information regarding the why the incident occurred \\
    \bottomrule
    \end{tabular}
} 
\end{table*}

\begin{table*}[!h]
\centering
\footnotesize
\caption{Codebook for Q.32 about the measures taken to prevent future fraud (vigilance).}
\label{tab:template}
\renewcommand{\arraystretch}{1.25}
\resizebox{\textwidth}{!}{%
    \begin{tabular}{p{0.3\textwidth}|p{0.7\textwidth}}
    \toprule 
    \rowcolor{gray!50}
    \multicolumn{2}{l}{Q.32: Measures to prevent future fraud $\rightarrow$ Vigilance} \\
    \midrule
    \textbf{Code (\# occurrence) [exclusive]}   & \textbf{Code Definition}     \\
    \hline
    unspecified (28) & Did not specify the security measures of type \quotes{vigilance} undertaken post the fraudulent experience \\
    \hline 
    cautious\_on\_websites (25)  & Exercised additional caution while using websites \\
    \hline 
    stop\_saving\_card\_in\_accounts (13) & Stopped saving card details in any digital account \\
    \hline 
    cautious\_card\_physical (8) & Took additional caution in securing card physically including avoiding handing over card when out of sight \\
    \hline 
    caution\_card\_pos\_systems (7) & Exercised additional caution in using card in point-of-sale systems \\
    \hline 
    cautious\_card\_use\_general (6) &	Exercised additional caution in general use \\
    \hline 
    use\_trusted\_financial\_services (6) &		Switched to trusted financial services with better security features like paypal, apple pay, etc. for certain transactions \\
    \hline 
    switch\_usage\_debit\_to\_credit (5) &		Switched usage from debit card to credit card \\
    \hline 
    cautious\_skimming\_device (4) & Exercised additional caution in ATM and point-of-sale systems to identify skimming device \\
    \hline 
    limit\_online\_transactions (3) &		Reduced online transactions \\
    \hline 
    stop\_using\_impacted\_service (3) &		Stopped using the service which caused the fraudulent incident \\
    \hline 
    caution\_card\_atm (2) &		Exercised caution when conducting transactions at the ATM including isolating themselves, checking for skimming device, etc. \\
    \hline 
    use\_one\_time\_cards\_for\_online (2) &	Utilized one-time card(s) referred as burner/virtual cards for online transactions \\
    \hline 
    cautious\_giving\_card\_info (1) &	Exercised additional caution when providing card information \\
    \hline 
    check\_transaction\_correctness (1) &	Exercised additional caution to ensure transaction details are correct, such as checking recipient, amount etc. \\
    \hline 
    diversify\_funds\_multiple\_accts (1) &		Diversified financial portfolio to include multiple bank or card issuers instead of one to reduce the risk/impact \\
    \hline 
    exercise\_caution\_while\_transacting (1) &		Exercised additional caution while conducting any financial transaction online/offline, or any mode of transaction \\
    \hline 
    limit\_online\_shopping (1) &		Reduced online transaction, specifically online shopping \\
    \hline 
    logout\_websites (1) &		Ensured Logging out of accounts after every usage to ensure that the accounts are not taken-over/misused \\
    \hline 
    not\_always\_work (1) &		When choosing not to save card info it doesn't work as expected (still save card details) \\
    \hline 
    require\_photo\_id (1) &		Opted for verification through photo ID for every transaction \\
    \hline 
    stop\_response\_suspicious\_msg (1) &		Stopped responding to any suspicious message that might lead to information disclosure \\
    \hline 
    urge\_family\_take\_same\_measures (1) &		Urged family members to take same measures (as the victims) to safeguard their accounts \\
    \hline 
    use\_giftcard\_on\_questionable\_websites (1) &		Used gift card when conducting transactions of less popular, questionable, or small business websites \\
    \hline 
    use\_protected\_wallet (1)	& Used protected wallets to safeguard card information from being stolen\\
    \bottomrule
    \end{tabular}
} 
\end{table*}

\begin{table*}[!h]
\centering
\footnotesize
\caption{Codebook for Q.32 about the measures taken to prevent future fraud (setup alerts).}
\label{tab:template}
\renewcommand{\arraystretch}{1.25}
\resizebox{\textwidth}{!}{%
    \begin{tabular}{p{0.3\textwidth}|p{0.7\textwidth}}
    \toprule 
    \rowcolor{gray!50}
    \multicolumn{2}{l}{Q.32: Measures to prevent future fraud $\rightarrow$ Setup alerts} \\
    \midrule
    \textbf{Code (\# occurrence) [non-exclusive]}   & \textbf{Code Definition}     \\
    \hline
    unspecified (20) & Did not specify the security measures of type \quotes{setup alerts} undertaken post the fraudulent experience \\
    \hline 
    unspecified\_alerts (10) &		Setup alerts. Unspecified type of alert configured \\
    \hline 
    any\_transaction\_alert (9) &	Setup alerts for all transactions \\
    \hline 
    sms\_alert (7) &		Setup alerts through SMS \\
    \hline 
    suspicious\_transaction\_alert (7) &	Setup alerts for suspicious transactions \\
    \hline 
    large\_transaction\_alert (6) &		Setup alerts for large transactions (this also includes transactions above a certain limit set by the user) \\
    \hline 
    app\_notifications (3) &	Setup alerts through app, e.g. push notifications \\
    \hline 
    email\_alert (3) &		Setup alerts through email \\
    \hline 
    enable\_notifications (3) &		Setup alerts for the bank or card issuer of the impacted card only \\
    \hline 
    all\_accounts (2) &		Setup alerts for all bank or card issuer accounts \\
    \hline 
    adjust\_minimum\_balance\_alert (1) &		Adjusted minimal balance for the corresponding minimum balance alert \\
    \hline 
    all\_alerts (1) &		Setup alerts for all activities on  the bank or card issuer account \\
    \hline 
    amount\_transaction\_alert (1) &		Setup alerts transactions of a certain amount and above \\
    \hline 
    check\_phone\_allow\_alerts (1) &		Verified the phone didn't suppress alerts configured on the bank or card issuer account \\
    \hline 
    credit\_report\_alert (1) &		Setup alerts for credit reports \\
    \hline 
    daily\_account\_activity\_alert (1) &		Setup alerts for daily activities on the bank or card issuer accounts \\
    \hline 
    multiple\_channels\_alerts (1) &		Setup alerts in multiple channels, e.g. phone, SMS, email etc. \\
    \hline 
    new\_purchase\_alert (1) &		Setup alerts for any new purchases \\
    \hline 
    security\_alerts (1) &		Setup alerts for security activities on the bank or card issuer \\
    \bottomrule
    \end{tabular}
} 
\end{table*}

\begin{table*}[!h]
\centering
\footnotesize
\caption{Codebook for Q.32 about the measures taken to prevent future fraud (regular checks).}
\label{tab:template}
\renewcommand{\arraystretch}{1.25}
\resizebox{\textwidth}{!}{%
    \begin{tabular}{p{0.3\textwidth}|p{0.7\textwidth}}
    \toprule 
    \rowcolor{gray!50}
    \multicolumn{2}{l}{Q.32: Measures to prevent future fraud $\rightarrow$ Regular checks} \\
    \midrule
    \textbf{Code (\# occurrence) [non-exclusive]}   & \textbf{Code Definition}     \\
    \hline
    review\_acct\_or\_card\_or\_statement (82) &	Review the bank account or card statement \\
    \hline 
    unspecified (27) & Did not specify the security measures of type \quotes{regular checks} undertaken post the fraudulent experience \\ 
    \hline 
    daily	(19) &		Review the bank account or card statement on a daily basis \\
    \hline 
    frequently (15) &		Review the bank account or card statement frequently, without mentioning the periodicity\\
    \hline 
	more\_frequently (8) &		Review the bank account or card statement more frequently as compared to before the fraudulent incident \\
    \hline 
	every\_few\_days (7) &		Review the bank account or card statement every few days \\
    \hline 
	app (4) & Review the bank account or card statement on the bank or card issuer's application \\
    \hline 
	few\_times\_a\_day (4) &		Review the bank account or card statement multiple times a day \\
    \hline 
	weekly (4) &		Review the bank account or card statement on a weekly basis \\
    \hline 
	few\_times\_a\_week (2) &		Review the bank account or card statement multiple times a week \\
	\hline
    website (2) &		Review the bank account or card statement on the bank or card issuer's website \\
	\hline
    before\_making\_payment (1) &  Review the bank account or card statement before making any payment \\
	\hline
    constantly (1) &		Review the bank account or card statement in a constant basis (without mentioning the periodicity) \\
	\hline
    few\_times\_a\_month (1) &		Review the bank account or card statement few times a month \\
	\hline
    immediately (1) &		Review the bank account or card statement immediately \\
	\hline
    monthly (2) &		Review their bank account or card statement in a monthly basis \\
	\hline
    multiple\_time\_a\_week (1) &		Review the bank account or card statement multiple times a week \\
	\hline
    phone (1) &		Review the bank account or card statement on their phone \\
    \bottomrule
    \end{tabular}
} 
\end{table*}

\begin{table*}[!h]
\centering
\footnotesize
\caption{Codebook for Q.32 about the measures taken to prevent future fraud (security measures).}
\label{tab:template}
\renewcommand{\arraystretch}{1.25}
\resizebox{\textwidth}{!}{%
    \begin{tabular}{p{0.3\textwidth}|p{0.7\textwidth}}
    \toprule 
    \rowcolor{gray!50}
    \multicolumn{2}{l}{Q.32: Measures to prevent future fraud $\rightarrow$ Security measures} \\
    \midrule
    \textbf{Code (\# occurrence) [non-exclusive]}   & \textbf{Code Definition}     \\
    \hline
    unspecified (30) & Did not specify the security measures of type \quotes{security measures} undertaken post the fraudulent experience \\
    \hline 
	2fa (7) &		Used software-based two-factor-authentication service for their bank or card issuer's accounts \\
    \hline 
	apply\_security\_measures (6) &		Utilized security measures in general, without mentioning what form of security measure were undertaken and applied \\
    \hline 
	reset\_password (4) &		Reset the password associated with their affected bank or card issuer's accounts \\
    \hline 
	change\_card (3) &		Changed the credit/debit card \\ 
    \hline 
	lock\_unused\_card (4) &		Locked the credit/debit card when not in use \\
    \hline 
	all\_accounts (3) &		Took similar security measures for all the bank or card issue's accounts (not only the affected one) \\
    \hline 
	lock\_card (2) &	 Locked the credit/debit card  \\
    \hline 
    use\_strong\_password (2) &		Configured strong password for the bank or card issuer's accounts \\
    \hline 
	cautious\_on\_websites (1) &		Exercised additional caution while using websites \\
    \hline 
	esignature\_all\_transactions (1) & Setup e-signature on all transactions \\ 
    \hline 
	extra\_security\_measures (1) &		Utilized additional security measures as compared to before the fraud incident \\
    \hline 
	freeze\_card (1) &		Froze the impacted card after the fraud incident \\
    \hline 
    freeze\_lost\_card (1) &		Froze the credit/debit card when losing the card \\
    \hline 
	lock\_card\_after\_suspicious\_transaction (1) &		Lock the credit/debit card when not in use \\
    \hline 
	logout\_websites (1) &		Logging out of the accounts after every usage to ensure the accounts is not taken-over/misused \\
    \hline 
	reset\_pin (1) &		Reset the PIN associated with the credit/debit card \\
    \hline 
	without\_card\_or\_acct\_freeze (1) &	Enabled alternate measure without freezing the cards \\
    \bottomrule
    \end{tabular}
} 
\end{table*}

\begin{table*}[!h]
\centering
\footnotesize
\caption{Codebook for Q.32 about the measures taken to prevent future fraud (security measures from third party).}
\label{tab:template}
\renewcommand{\arraystretch}{1.25}
\resizebox{\textwidth}{!}{%
    \begin{tabular}{p{0.3\textwidth}|p{0.7\textwidth}}
    \toprule 
    \rowcolor{gray!50}
    \multicolumn{2}{l}{Q.32: Measures to prevent future fraud $\rightarrow$ Security measures from third party} \\
    \midrule
    \textbf{Code (\# occurrence) [non-exclusive]}   & \textbf{Code Definition}     \\
    \hline 
    unspecified (12) & Did not specify the security measures of type \quotes{third party} undertaken post the fraudulent experience\\
    \hline 
    credit\_monitoring\_svc (6) &	 Started using credit monitoring service \\
    \hline 
    identity\_protection\_service (4) &		Started using identity protection service to secure digital identities \\
    \hline
    2fa (1) &	Started using software based two-factor-authentication service for digital accounts\\
    \hline 
    any\_transaction\_alert (1) &	Configured alerts for any transactions on third party services \\
    \hline 
    2fa\_hw (1) &		Started using hardware token-based two-factor-authentication service for digital accounts \\
    \hline 
    credit\_report (1) &	Started reviewing the credit report \\
    \hline 
    fraud\_app (1) &	Started using fraud app \\
    \hline 
    protection\_sw (1) &		Started using software to protect laptop, phone, or other devices \\ 
    \hline 
    reset\_password (1) &		Reset the password of affected account(s) \\
    \hline 
    reset\_phone (1) &		Performed factory reset for the affected phone \\
    \hline 
    use\_password\_manager (1) &		Started using password manager to secure their passwords \\
    \bottomrule
    \end{tabular}
} 
\end{table*}

\begin{table*}[!h]
\centering
\footnotesize
\caption{Codebook for Q.32 about the measures taken to prevent future fraud (change bank or card issuer).}
\label{tab:template}
\renewcommand{\arraystretch}{1.25}
\resizebox{\textwidth}{!}{%
    \begin{tabular}{p{0.3\textwidth}|p{0.7\textwidth}}
    \toprule 
    \rowcolor{gray!50}
    \multicolumn{2}{l}{Q.32: Measures to prevent future fraud $\rightarrow$ Change bank or card issuer} \\
    \midrule
    \textbf{Code (\# occurrence) [non-exclusive]}   & \textbf{Code Definition}     \\
    \hline
	change\_bank (8) &		Changed the bank \\
    \hline 
    unspecified (2) &		Did not specify the security measures of type \quotes{change bank or card issuer} undertaken post the fraudulent experience \\
    \hline 
    switch\_usage\_debit\_to\_credit (2) &		Switched to using credit card instead of debit card owing to better security measures \\
    \hline 
    change\_bank\_card (1) &		Replaced the affected card with a new card with the same bank or card issuer \\
    \hline 
    diversify\_funds\_multiple\_banks (1) &	Diversified financial portfolio to include multiple bank or card issuers instead of one to reduce the risk/impact \\
    \bottomrule
    \end{tabular}
} 
\end{table*}

\end{document}